\documentclass[11pt,a4paper]{article}
\usepackage{jcappub}
\usepackage{mathtools}
\usepackage{graphicx}
\usepackage{color}
\usepackage{multirow}
\usepackage{amsmath}
\usepackage[amssymb]{SIunits}
\usepackage{cancel}
\usepackage{here}

\title{
  Big-bang nucleosynthesis with sub-GeV massive decaying
  particles
}

\author[a,b]{Masahiro Kawasaki}
\author[c,d,b]{Kazunori Kohri}
\author[e,b]{Takeo Moroi}
\author[a,b]{Kai Murai}
\author[f,g,b]{Hitoshi Murayama\thanks{Hamamatsu Professor}}
\affiliation[a]{ Institute for Cosmic Ray Research,
  The University of Tokyo, Kashiwa 277-8582, Japan}
\affiliation[b]{ Kavli Institute for the Physics and Mathematics of the
  Universe (WPI), University of Tokyo,
  Kashiwa 277-8583, Japan}
\affiliation[c]{ Theory Center, IPNS, KEK, Tsukuba 305-0801, Japan}
\affiliation[d]{ The Graduate University of Advanced Studies (Sokendai), Tsukuba 305-0801, Japan}
\affiliation[e]{Department of Physics, The University of Tokyo,
Tokyo 113-0033, Japan}
\affiliation[f]{ Department of Physics, University of California, Berkeley, CA 94720, USA}
\affiliation[g]{ Ernest Orlando Lawrence Berkeley National Laboratory, University of California, Berkeley, CA 94720, USA}

\date{\today}
                

\abstract{

  We consider the effects of the injections of energetic photon and
  electron (or positron) on the big-bang nucleosynthesis.  We study
  the photodissociation of light elements in the early Universe 
  paying particular attention to the case that the injection energy is sub-GeV 
  and derive upper bounds on the primordial abundances of
  the massive decaying particle as a function of its lifetime.  
  We also discuss a solution of the $^7$Li problem in this
  framework.

}

\begin{document}

\begin{flushright}
%  IPMU18-xxxx,
  KEK-Cosmo-254, KEK-TH-2214
\end{flushright}
\maketitle

%%%%%%%%%%%%%%%%%%%%%%%%%%%%%%%%%%%%%%%%%%%%%%%%%%%%%%%%%%%%%%%%%%%%%%
\section{Introduction}
%%%%%%%%%%%%%%%%%%%%%%%%%%%%%%%%%%%%%%%%%%%%%%%%%%%%%%%%%%%%%%%%%%%%%%

Big-bang nucleosynthesis (BBN) provides a powerful probe of the
thermal history of the Universe.  In the standard BBN (SBBN) scenario,
the light elements (i.e., D, $^3$He, $^4$He, and so on) are
synthesized at the cosmic time of $t\gtrsim 1\ {\rm sec}$
(corresponding to the cosmic temperature of $T\lesssim 1\ {\rm MeV}$)
at which the typical energy of the background photon becomes much
lower than the binding energies of nuclei.  Because the predictions of
the SBBN are more or less in good agreements with the observations of
the primordial abundances of light elements \cite{Tanabashi:2018oca},
beyond-the-standard-model (BSM) physics which alters the thermal
history after $\sim 1\ {\rm sec}$ may be constrained or excluded in
order not to spoil the success of the SBBN.

One important class of BSM models affecting the light element
abundances is that with long-lived particles.  In many BSM models,
there shows up a long-lived particle whose lifetime is longer than
$\sim 1\ {\rm sec}$.  If such a long-lived particle is somehow
produced in the early Universe, and also if it decays into
standard-model particles, its late-time decay causes hadronic and
electromagnetic showers after the BBN epoch.  Energetic particles in
the showers dissociate light elements, resulting in the change of
their abundances.  The effects of radiative decay of long-lived
particles on the light element abundances have been intensively
studied in
Refs.~\cite{photodisOLD,Kawasaki:1994af,Kawasaki:1994sc,Moroi:1995fs,Holtmann:1998gd,Jedamzik:1999di,Kawasaki:2000qr,Cyburt:2002uv,Jedamzik04a,Kawasaki:2004yh,Kawasaki:2004qu,Jedamzik:2006xz,Cyburt:2009pg,Kawasaki:2017bqm}
and references therein (see also Refs.~\cite{{Reno:1987qw,DimEsmHalSta,Kohri:2001jx,Jedamzik04a,Kawasaki:2004yh,Kawasaki:2004qu,Jedamzik:2006xz,Cyburt:2009pg,Kawasaki:2017bqm}} for hadronic decays).  Many of previous studies paid particular
attention to the case where the mass of the long-lived particle is
around or above the electroweak scale, which is motivated by the
relation between the BSM physics and the electroweak scale.  In
particular, in the thermal relic dark matter scenario, the pair
annihilation cross section of the dark matter particle is about
$1\ \rm {\rm pb}$, corresponding to the cross section obtained by the
exchange of particle with mass around the electroweak scale (assuming
that the coupling constant is sizable).  It motivates us to consider
BSM models for dark matter whose typical mass scale is around the
electroweak scale.

Recently, however, models with sub-GeV dark matter have been
attracting much attention
\cite{Hochberg:2014dra,Hochberg:2014kqa,Lee:2015gsa,Hochberg:2015vrg,Berlin:2018tvf}.
In such models, the masses of particles other than the dark matter
particle are also often sub-GeV.  Importantly, in certain models
\cite{Hochberg:2018vdo}, some of the particles in the model become
long-lived and may have lifetime longer than $\sim 1\ {\rm sec}$.  If
so, such long-lived particles may affect the light element abundances.
Effects of sub-GeV long-lived particles on the BBN significantly
differs from those of heavier ones.
\begin{itemize}
\item If the mass of the long-lived particle is around the electroweak
  scale (or higher), significant amount of hadrons are expected to be
  produced by the decay, causing hadronic shower in the thermal plasma
  in the early
  Universe~\cite{{Reno:1987qw,DimEsmHalSta,Kohri:2001jx,Jedamzik04a,Kawasaki:2004yh,Kawasaki:2004qu,Jedamzik:2006xz,Cyburt:2009pg,Kawasaki:2017bqm}}.
  Then, the hadrodissociation process often gives stringent
  constraints on the model~\cite{{Kawasaki:2017bqm}} (even though
  electromagnetic shower also occurs).  For the case of sub-GeV
  long-lived particle, on the contrary, productions of hadrons at the
  time of the decay are kinematically suppressed (or forbidden).  In
  such a case, effects of photodissociation processes of light
  elements become more important.
\item With the injection of energetic photons (or electrons and positrons) into
  thermal plasma, electromagnetic showers are induced.  The photon
  spectrum in the showers is (approximately) determined by the total
  amount of energy injection if the energy of primary photons is large
  enough \cite{Kawasaki:1994sc,Moroi:1995fs}.  If the energy of the primary
  photon is sub-GeV, on the contrary, such a universal behavior of the
  spectrum is lost and the shape of the spectrum becomes sensitive to
  the energy of the injected photons
  \cite{Poulin:2015woa,Poulin:2015opa,Forestell:2018txr} (See also
  \cite{Acharya:2019uba}).  Thus, for the study of the BBN
  constraints on the sub-GeV long-lived particles, a careful
  calculation of the photon spectrum in the electromagnetic shower is
  needed.
\end{itemize}
Thus, a detailed study about the effects of sub-GeV long-lived
particles on the BBN predictions should be necessary.  

Motivated by these observations, we study the light element abundances
in models with sub-GeV long-lived particles.  We concentrate on the
case where the unstable particle decays into photon or electron and
positron, assuming that the decay processes into other particles (in
particular, hadrons) are kinematically suppressed or forbidden.

The organization of this paper is as follows.  In Section
\ref{sec:obs}, we summarize the observational abundances of light
elements.  In Section \ref{sec:shower}, we discuss our treatment of
the electromagnetic shower induced by the injection of the
electromagnetic particles into the thermal bath in the early Universe.
In Section \ref{sec:entropyprod}, we discuss the entropy production
which also plays important role in constraining the long-lived
particles.  The constraints on the primordial abundance of long-lived
particles are given in Section \ref{sec:results}.  A possibility to
solve the $^7$Li problem using long-lived particles are discussed in
Section \ref{sec:liprob}.  Section \ref{sec:conclusions} is devoted to
conclusions and discussion.
    
In this paper, we adopt natural units, $c = \hbar = 1$, and 
$n_i$ indicates the number density of species ``i.''  In addition,
yield variable is defined as
\begin{align}
  Y_i \equiv \frac{n_i}{s},
\end{align}
with $s$ being the entropy density.

%%%%%%%%%%%%%%%%%%%%%%%%%%%%%%%%%%%%%%%%%%%%%%%%%%%%%
\section{Observational Abundances of Light Elements}
\label{sec:obs}
\setcounter{equation}{0}
%%%%%%%%%%%%%%%%%%%%%%%%%%%%%%%%%%%%%%%%%%%%%%%%%%%%\

In this section we summarize observational constraints on the primordial abundances of light elements (D, $^3$He, $^4$He and $^7$Li).  
The errors are written at 68$\%$ C.L. unless otherwise
noted. The subscript $p$ denotes the primordial value.

\begin{itemize}
    \item D\\
    The deuterium abundance has been precisely measured by observing absorption of QSO lights due to damped Lyman-$\alpha$ systems.
    We adopt
    \begin{equation}
        (\text{D}/\text{H})_p = (2.545 \pm 0.025)\times 10^{-5},
        \label{eq:D_obs}
    \end{equation}
    from measurements of 13 damped Lyman-$\alpha$ systems~\cite{Zavarygin:2018dbk}.  This
    abundance is consistent with another recent value $(\text{D}/\text{D})_p = (2.527\pm
    0.030)\times 10^{-5}$ reported by Cooke \textit{et
      al.}~\cite{Cooke:2017cwo}.
    \item $^3$He\\ We adopt the observed $^3$He/D as an upper bound on the
      primordial value~\cite{Sigl:1995kk}.  From $^3$He and D
      abundances observed in protosolar clouds~\cite{Geiss:2003}, we adopt
    \begin{equation}
        ({\rm ^3He}/{\rm D})_p < 0.83 + 0.27.
        \label{eq:obs_const_He3D}
    \end{equation}
    \item $^4$He\\ The primordial abundance of $^4$He is determined by
      measurement of recombination lines from extra-galactic HII
      regions.  Izotov \textit{et al.}~\cite{Izotov:2014fga} obtained
      $Y_p = 0.2551\pm 0.0022$ from the observation of 45
      extragalactic HII regions.  Aver, Olive and
      Skillman~\cite{Aver:2015iza} reanalyzed the data of
      Ref.~\cite{Izotov:2014fga} and obtained $Y_p= 0.2449 \pm 0.0040$
      which is inconsistent with the value given in
      \cite{Izotov:2014fga}.  More recently Fern\'andez \textit{et
        al.}~\cite{Fernandez:2018} reported $Y_p=0.245\pm 0.007$ using
      27 HII regions selected SDSS.  Valerdi \textit{et
        al.}~\cite{Valerdi:2019} obtained $Y_p = 0.2451\pm 0.0026$
      from the observation of the HII region in NGC 246.  These recent
      measurements are in good agreement with result of
      \cite{Aver:2015iza}.  Therefore, in this paper we adopt the
      value obtained by Aver, Olive and Skillman~\cite{Aver:2015iza},
    \begin{equation}
       Y_p = 0.2449 \pm 0.0040.
       \label{eq:obs_const_He4_AOS}
    \end{equation}
    \item $^7$Li\\ Observations of $^7$Li abundances in atmospheres of
      metal-poor halo stars show an almost constant value called
      "Spite plateau" which has been considered as primordial.
      Bonifacio et al~\cite{Bonifacio:2006au} reported the $^7$Li
      abundance as
    \begin{eqnarray}
      {\rm Log}_{10}\left({^7{\rm Li}}/{\rm H}\right)_p = -9.900\pm 0.090.
      \label{eq:Li7}
    \end{eqnarray}
    However, the above abundance is about three times smaller than
    that predicted in the standard BBN. This discrepancy between the
    BBN prediction and observation is called ``$^7$Li problem.''
    However, recent observations of extremely metal-poor stars show
    abundance much smaller than the Spite plateau value
    \cite{Aoki:2009ce}.  If we take such recent observations
    seriously, some unknown physical processes should change the
    $^7$Li abundances during or after the BBN.  Thus, it may be still
    premature to regard the abundance given in Eq.\ \eqref{eq:Li7} as
    a primordial abundance of $^7$Li.  In the following, we will
    derive upper bounds on the primordial abundance of the unstable
    particle $X$. In such an analysis, we try to be conservative so
    that we do not use $^7$Li abundances for the study of the bound.
    In addition, we will also discuss the implication of the
    long-lived particle on the $^7$Li problem.  We will show that MeV
    decaying particles may give a solution to the $^7$Li problem if
    the plateau abundance \eqref{eq:Li7} is primordial
    \cite{Bonifacio:2006au} (and lower abundances in extremely metal-poor
    stars is realized by another process,
    e.g. \cite{Kusakabe:2019li}).
    
\end{itemize}

Here we also mention other observational constraints used in the present paper.
The baryon to photon ratio is taken to be
\begin{align}
  \eta = (6.13 \pm 0.04) \times 10^{-10},
  \label{eq:etaobs}
\end{align}
which is based on the density parameter for
baryon  $\Omega_{\rm B} h^2 = 0.02237^{+0.00015}_{-0.00015}$
(68$\%$ C.L.) for the TT,TE,EE+lowE+lensing analysis by the Planck
collaboration, where $h$ is the Hubble constant in units of $100\ {\rm
  km/sec/Mpc}$~\cite{Aghanim:2018eyx}. 

The Planck collaboration also reported constraints on the
effective number of neutrino species. We adopt the following value in the same
case as that of $\eta$ by the TT,TE,EE+lowE (+lensing) analysis,
\begin{align}
  N_{\nu}^{\rm eff}  = 2.92 ^{+0.36}_{-0.37},
  \label{eq:NeffnuObs}
\end{align}
at 95$\%$ C.L. ~\cite{Aghanim:2018eyx}. That means $N_{\nu}^{\rm eff}$
can ranges 2.55 -- 3.28 at 95$\%$ C.L. where the lower limit
$N_{\nu}^{\rm eff} > N_{\nu,{\rm low}}^{\rm eff} = 2.55$ is important
to constrain late-time injections of electromagnetic energy produced by massive particles decaying at
$T \ll 1$~MeV~\cite{Kawasaki:1999na}.

%%%%%%%%%%%%%%%%%%%%%%%%%%%%%%%%%%%%%%%%%%%%%%%%%%%%%%%%%%%%%%%%%%%%%%
\section{Electromagnetic Shower}
\label{sec:shower}
%%%%%%%%%%%%%%%%%%%%%%%%%%%%%%%%%%%%%%%%%%%%%%%%%%%%%%%%%%%%%%%%%%%%%%

In this section, we discuss how we treat the electromagnetic shower
induced by the decaying particle (which we call $X$).  We follow the
procedure explained in Refs.~\cite{Kawasaki:1994af, Kawasaki:1994sc}.

For the cosmic temperature of our interest, relevant scattering
processes induced by energetic photons are as follows.
\begin{itemize}
\item Double photon pair creation: $\gamma+\gamma_{\rm BG}\rightarrow
  e^++e^-$.
\item Photon photon scattering: $\gamma+\gamma_{\rm BG}\rightarrow
  \gamma+\gamma$.
\item Pair creation in nuclei: $\gamma+N_{\rm BG}\rightarrow
  \gamma+e^++e^-$.
\item Compton Scattering: $\gamma+e^-_{\rm BG}\rightarrow
  \gamma+e^-$.
\end{itemize}
In addition, energetic electron (as well as positron) loses its energy
via
\begin{itemize}
\item Inverse Compton scattering: $e^\pm+\gamma_{\rm BG}\rightarrow
  e^\pm + \gamma$.
\end{itemize}
Here, the subscript ``BG'' is for particles in the background and $N_{\rm BG}$ denotes the background nuclei.

We concentrate on the case where the scattering rates of photons and
electrons are larger than the expansion rate of the Universe.  Then,
neglecting the effects of the cosmic expansion, the Boltzmann equations
for the distribution function of photons (denoted as $f_{\gamma}$) and
that of electrons plus positrons (denoted as $f_e$) are denoted as
\begin{align}
  \frac{\partial f_{\gamma}(E_\gamma)}{\partial t}
  =&
  \left[ \frac{\partial f_{\gamma}(E_\gamma)}{\partial t} \right]_{\rm DP}
  + \left[ \frac{\partial f_{\gamma}(E_\gamma)}{\partial t} \right]_{\rm PP}
  + \left[ \frac{\partial f_{\gamma}(E_\gamma)}{\partial t} \right]_{\rm PC}
  + \left[ \frac{\partial f_{\gamma}(E_\gamma)}{\partial t} \right]_{\rm CS}
  + \left[ \frac{\partial f_{\gamma}(E_\gamma)}{\partial t} \right]_{\rm IC}
  \nonumber \\
  &  + \left[ \frac{\partial f_{\gamma}(E_\gamma)}{\partial t} \right]_{\rm DE},
  \label{fdot_pho} \\
  \frac{\partial f_{e}(E_e)}{\partial t}
  =&
  \left[ \frac{\partial f_{e}(E_e)}{\partial t} \right]_{\rm DP}
  + \left[ \frac{\partial f_{e}(E_e)}{\partial t} \right]_{\rm PC}
  + \left[ \frac{\partial f_{e}(E_e)}{\partial t} \right]_{\rm CS}
  + \left[ \frac{\partial f_{e}(E_e)}{\partial t} \right]_{\rm IC}
  + \left[ \frac{\partial f_{e}(E_e)}{\partial t} \right]_{\rm DE},
  \label{fdot_ele}
\end{align}
where terms with the index DP (PP, PC, CS, IC, and DE) represents the
contribution from the double photon pair creation process
(photon-photon scattering, pair creation in nuclei, Compton
scattering, Inverse Compton scattering, and decay of the exotic
particle).  Explicit forms of individual terms are given in
Ref.~\cite{Kawasaki:1994sc}.

Hereafter, for simplicity, we concentrate on the case that the energy
of the particles produced by the decay is monochromatic (and is
denoted as $\epsilon_0$) except for the final-state radiation (FSR).
Then, the decay terms take a simple form:
\begin{itemize}
\item For the case of monochromatic photon injection from the decay,
  the decay terms are given by
  \begin{align}
    \left[ \frac{\partial f_{\gamma}(E_\gamma)}{\partial t} \right]_{\rm DE}
    =\, &
    \xi_\gamma \frac{n_X}{\tau_X} \delta (E_\gamma - \epsilon_0),
    \\
    \left[ \frac{\partial f_e(E_e)}{\partial t} \right]_{\rm DE}
    =\, & 
    0,
  \end{align}
  where $n_X$ is the number density of $X$, $\tau_X$ is the lifetime
  of $X$, and $\xi_\gamma$ is the number of photon emitted by the decay
  of one $X$ particle.
\item For the case where $X$ decays primarily into monochromatic
  electron and positron, energetic photons are also produced via FSR,
  as discussed in Ref.~\cite{Forestell:2018txr} (see also Appendix~\ref{sec:FSR} for a generic case).  Taking into account
  the effect of FSR, the decay terms are given by
  \begin{align}
    \left[ \frac{\partial f_{\gamma}(E_\gamma)}{\partial t} \right]_{\rm DE}
    = \, & 
    \xi_e
    \frac{n_X}{\tau_X \epsilon_0}
    \frac{\alpha}{2\pi}
    \frac{1+(1-x)^2}{x}
    \ln \left[ \frac{4 \epsilon_0^2 (1-x)}{m_e^2} \right]
    \theta \left( 1 - \frac{m_e^2}{4 \epsilon_0^2} - x \right), 
    \label{fgmmdot_ein}
    \\
    \left[ \frac{\partial f_e(E_e)}{\partial t} \right]_{\rm DE}
    = \, & 
    \xi_e
    \frac{n_X}{\tau_X} \delta (E_e - \epsilon_0),
  \end{align}
  where $\alpha$ is the fine structure constant, and $x\equiv
  E_\gamma/\epsilon_0$.  In addition, $\xi_e$ is the number of
  electrons and positrons produced by the decay of $X$.  (If $X$ decays
  into a $e^+e^-$ pair, for example, $\xi_e=2$.)
\end{itemize}

We numerically solve Eqs.~\eqref{fdot_pho} and \eqref{fdot_ele} to
obtain the distribution functions of photons and electrons.  Because the
time scale of the shower evolution is much shorter than the time scale
of the cosmic expansion, we solve the Boltzmann equations taking
$\dot{f}_\gamma(E_\gamma)=\dot{f}_e(E_e)=0$ (with the ``dot''
indicating the derivative with respect to time) for each $t$,
approximating $\dot{n}_X=0$.

\begin{figure}[t]
\begin{center}
\includegraphics[width=0.4\textwidth]{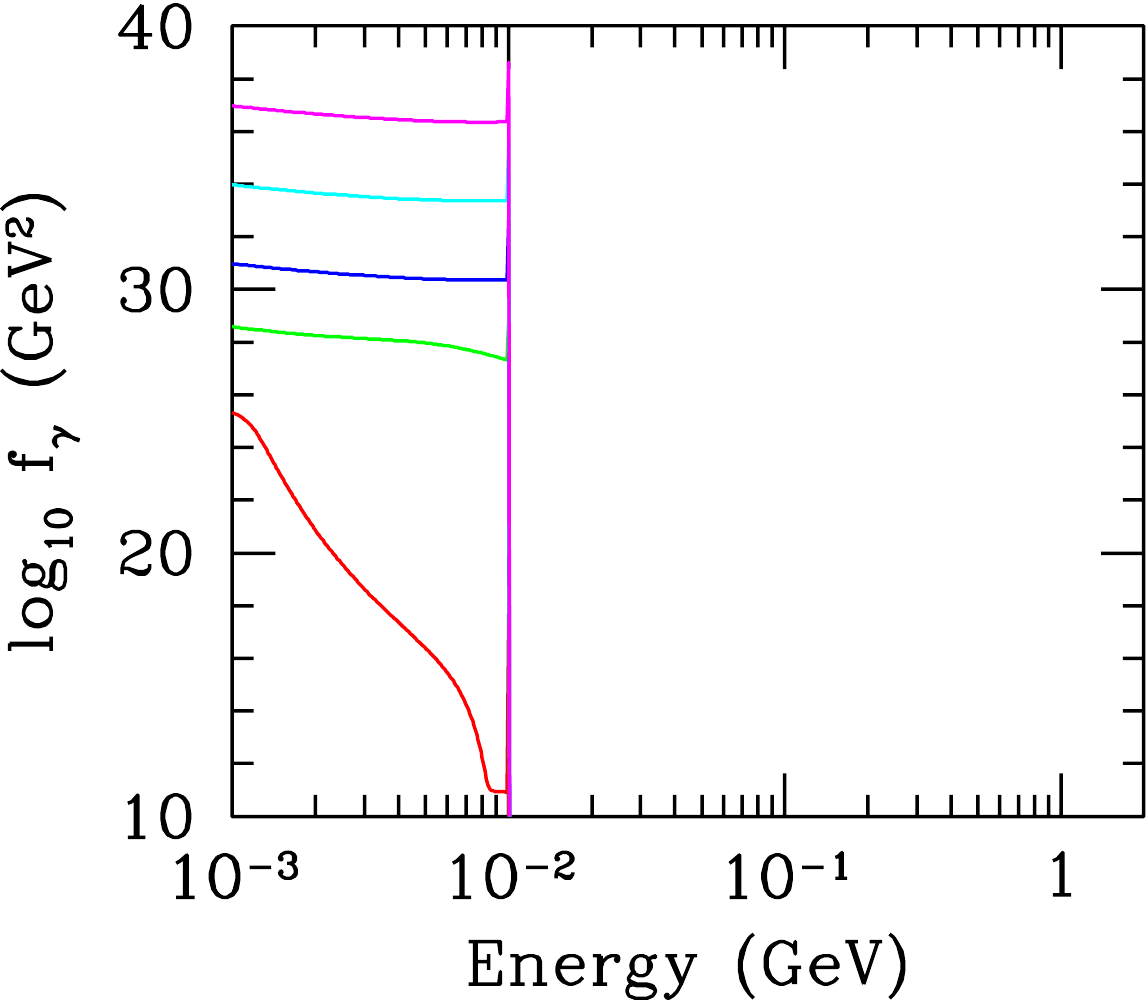}~~~~~~~
\includegraphics[width=0.4\textwidth]{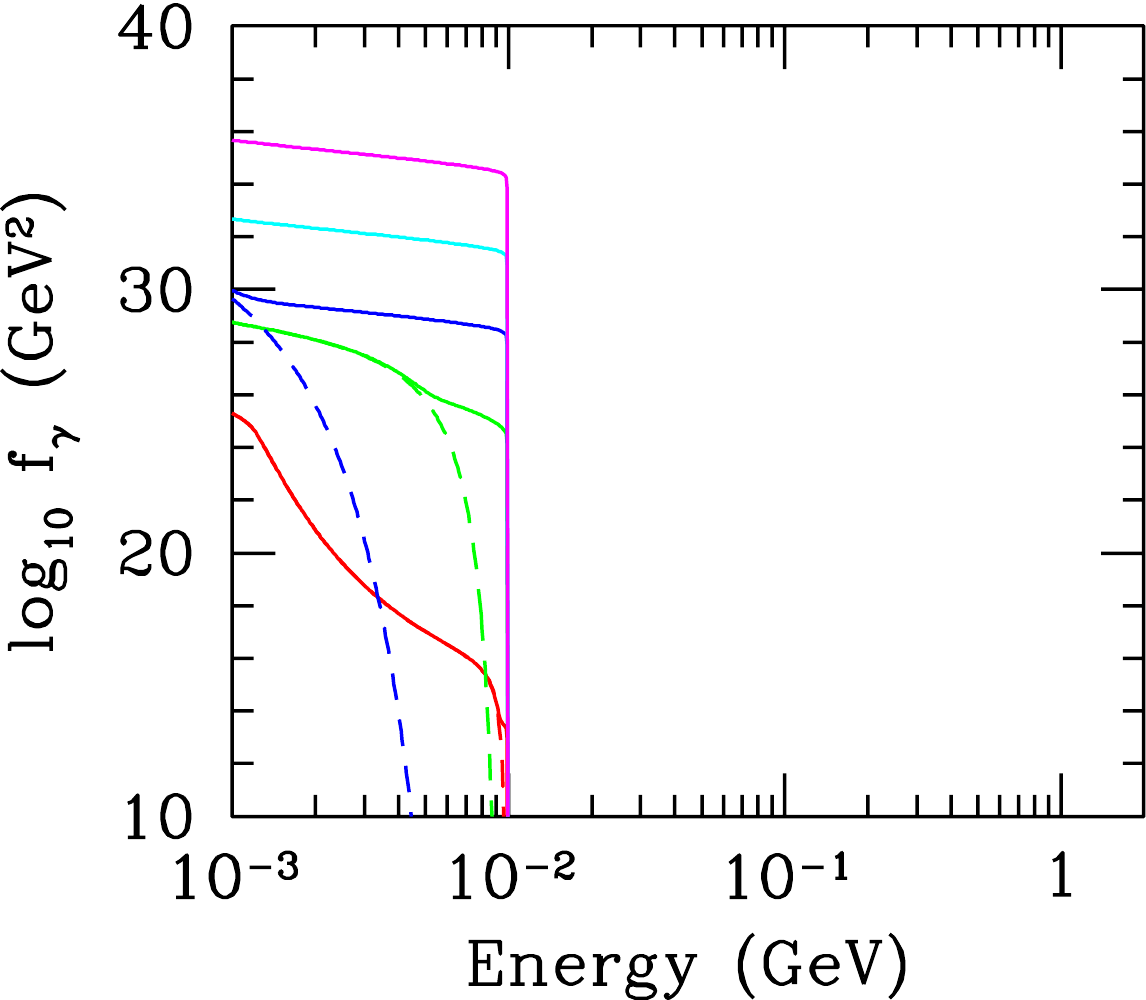}
\caption{Photon spectrum induced by monochromatic photon (left) and
  electron (right) injection, taking $\epsilon_0=10\ {\rm MeV}$.  The
  cosmic temperature is $10^{-5}$, $10^{-6}$, $10^{-7}$, $10^{-8}$,
  and $10^{-9}\ {\rm GeV}$, from below.  Here, we take $n_X=1\ {\rm
    GeV}^3$.  For the case of the monochromatic electron injection,
  the solid lines show the spectrum with the effect of FSR, while
  the dashed lines show the spectrum with neglecting the 
  effect of FSR (taking $[\dot{f}_\gamma]_{\rm DE}=0$, instead of
  Eq.\ \eqref{fgmmdot_ein}).}
\label{fig:spect0010}
\end{center}
\begin{center}
\includegraphics[width=0.4\textwidth]{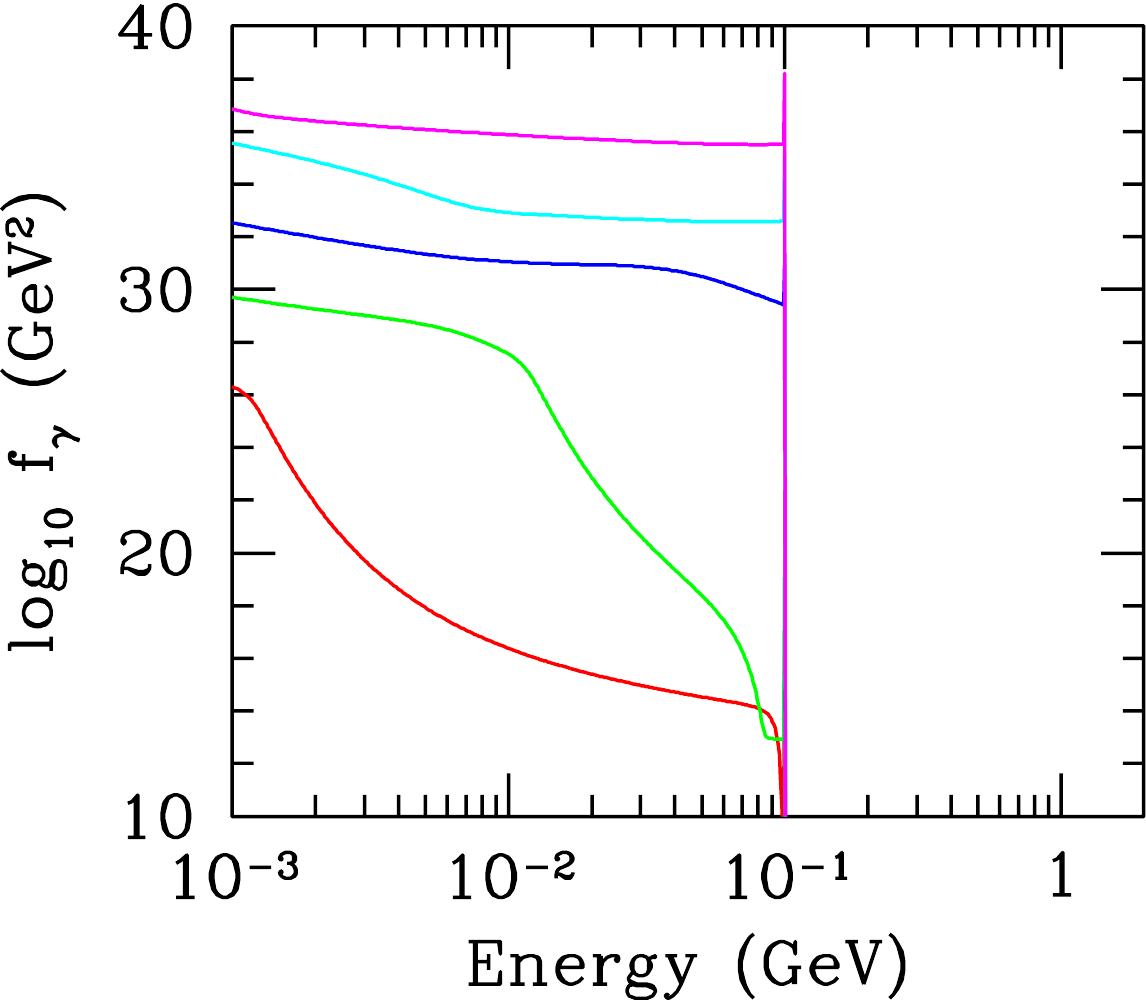}~~~~~~~
\includegraphics[width=0.4\textwidth]{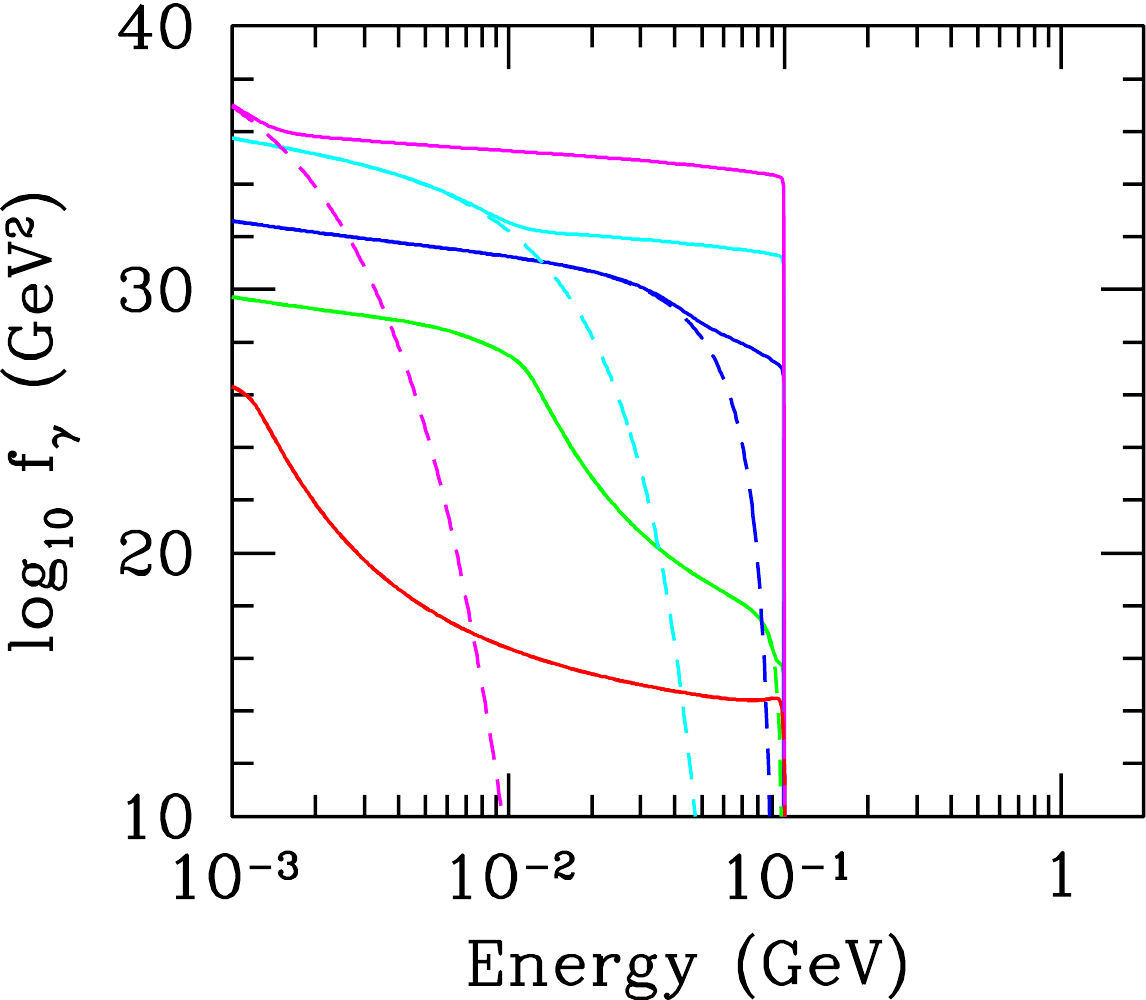}
\caption{Same as Fig.\ \ref{fig:spect0010}, except for $E_{\rm
    in}=100\ {\rm MeV}$.}
\label{fig:spect0100}
\end{center}

\end{figure}

\begin{figure}[t]
\begin{center}
\includegraphics[width=0.4\textwidth]{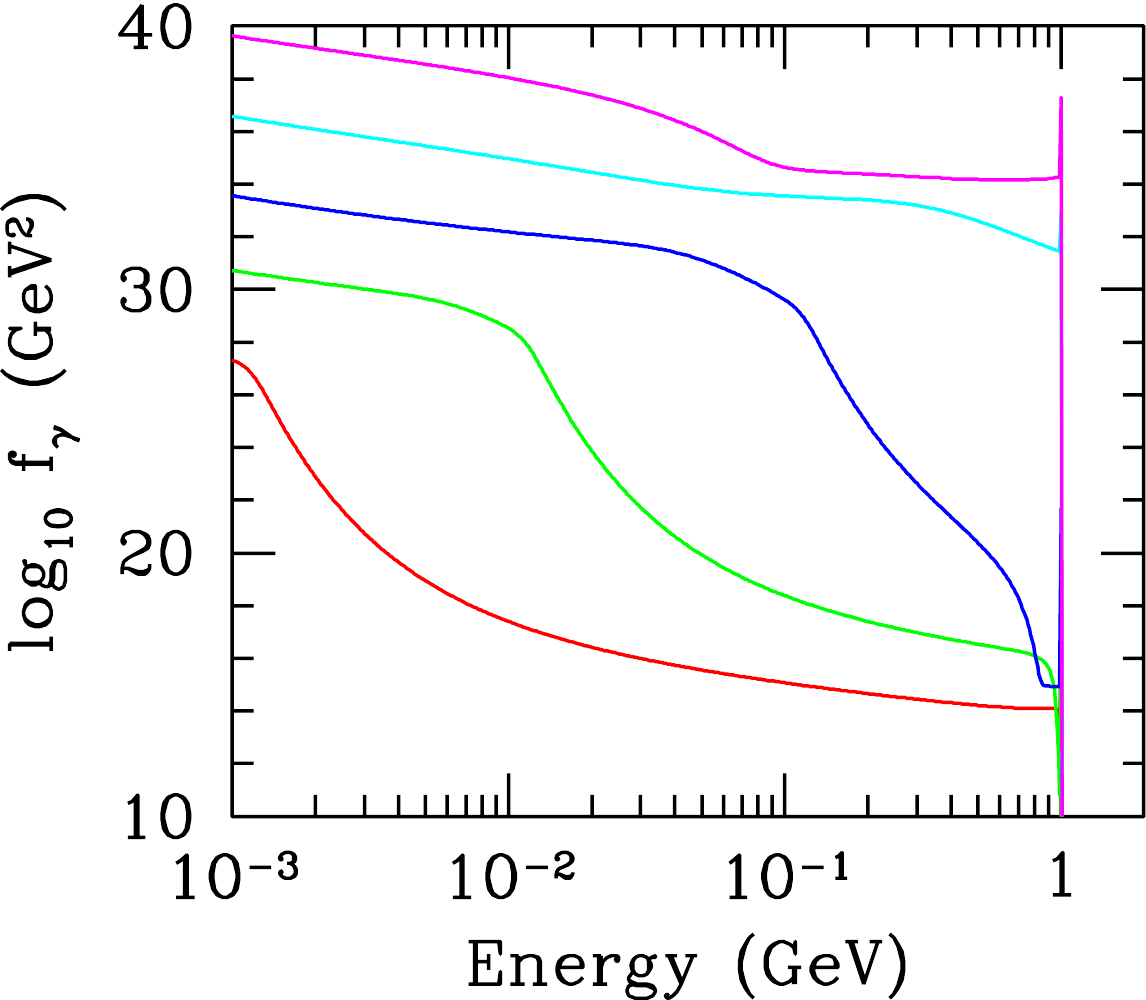}~~~~~~~
\includegraphics[width=0.4\textwidth]{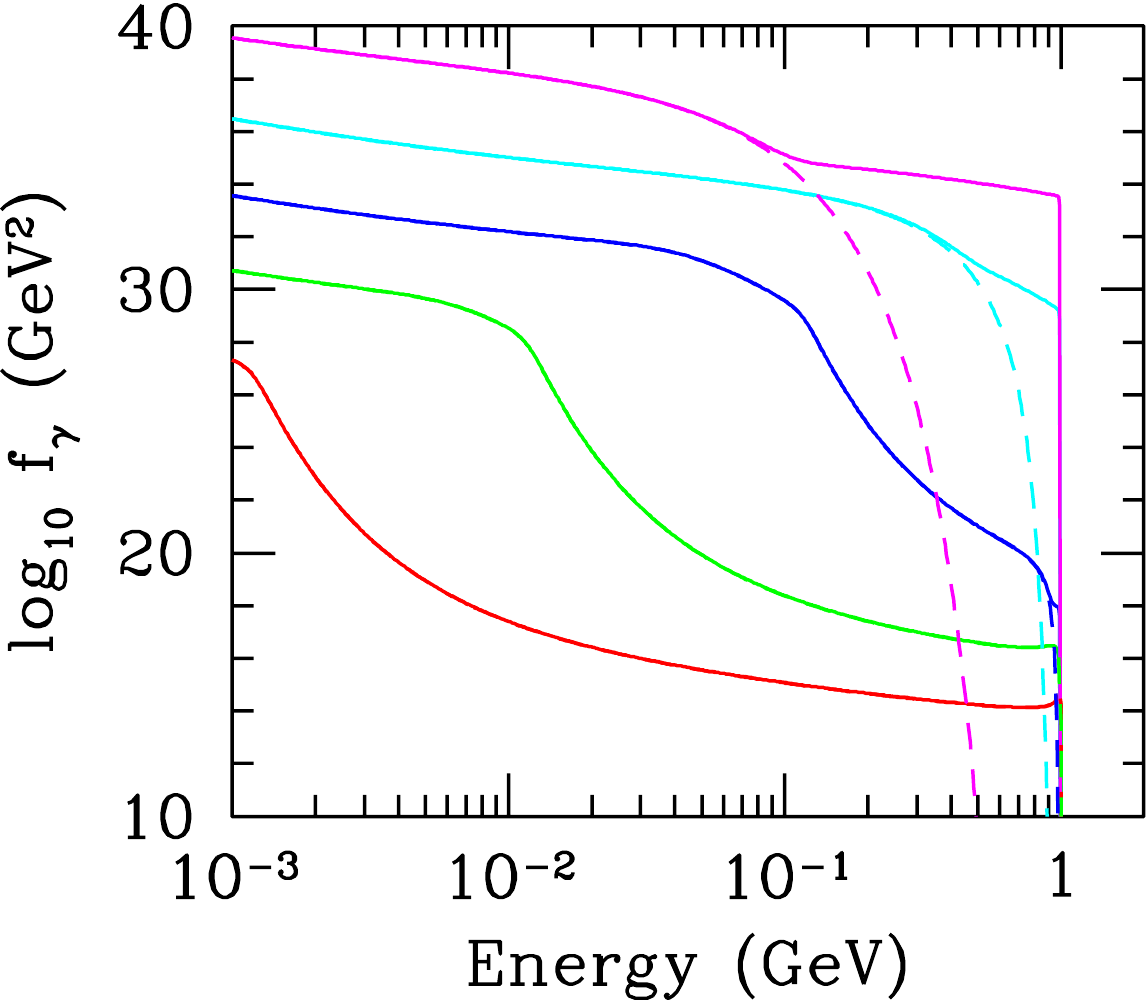}
\caption{Same as Fig.\ \ref{fig:spect0010}, except for $E_{\rm
    in}=1\ {\rm GeV}$.}
\label{fig:spect1000}
\end{center}
\end{figure}

In Figs.~\ref{fig:spect0010}, \ref{fig:spect0100}, and
\ref{fig:spect1000}, we show the photon spectrum induced by the
injections of monochromatic photons and electrons, taking $E_{\rm
  in}=10\ {\rm MeV}$, $100\ {\rm MeV}$, and $1\ {\rm GeV}$,
respectively.  The singular peaks at $E_\gamma=\epsilon_0$ are due to
the monochromatic injection of the photons.

We can see that, for the case of the electron injection, the FSR
photons significantly affect the photon spectrum in particular when
the cosmic temperature becomes low.  This is because, when $X$
primarily decays into a $e^+e^-$ pair, the electromagnetic shower is
initiated by the FSR photons as well as Inverse Compton scattering process $e^\pm+\gamma_{\rm BG}\rightarrow e^\pm\gamma$.  
When the electron becomes non-relativistic in the center-of-mass frame, the energy transfer of the final-state photon in the Inverse Compton process
becomes typically $\sim T$.  
Because we are interested in the photons
spectrum for $E_\gamma\gtrsim O(1)\ {\rm MeV}$ for the study of the
photodissociation processes, the Inverse Compton process becomes
unimportant in such a case.  The FSR photons are, on the contrary,
more energetic and can affect the photon spectrum relevant for the
photodissociation.  The FSR photons carry away only a fraction of the
energy (i.e., the mass of $X$).  Thus, the photon spectrum mainly initiated by the FSR photons is suppressed compared to the case of the energetic photon
injection.  This has significant implication for the BBN constraints on
unstable particles as we will discuss below.

%%%%%%%%%%%%%%%%%%%%%%%%%%%%%%%%%%%%%%%%%%%%%%%%%%%%%%%%%%%%%%%%%%%%%%

%%%%%%%%%%%%%%%%%%%%%%%%%%%%%%%%%%%%%%%%%%%%%%%%%%%%%%%%%%%%%%%%%%%%%%

For the case of the injection of monochromatic electron, we observe
deviation of our photon spectrum from those of
\cite{Forestell:2018txr}; for some choice of parameters, our photon
spectrum is smaller by a factor of $\sim 2$.  We expect that the
deviation originates from the difference of the treatment of the
effect of Compton scattering at small $y_e\equiv E_e T/m_e^2$.
The energy-loss rate of the electron in thermal bath is given by
\cite{Blumenthal:1970gc}
\begin{align}
  \dot{E}_e = -\frac{4}{3} \sigma_{\rm T} 
  \gamma_e^2
  \left[ 1 - \frac{63}{10}
    \frac{\gamma_e \langle \epsilon_\gamma^2 \rangle}
    {m_e \langle \epsilon_\gamma \rangle}
    + \cdots
  \right] \rho_\gamma,  
\end{align}
where $\langle\epsilon_\gamma\rangle$ and
$\langle\epsilon_\gamma^2\rangle$ are the mean energy and mean energy
squared of the background photons, respectively, $\sigma_{\rm T}\equiv
(8\pi/3) \alpha^2/m_e^2$, and $\gamma_e\equiv E_e/m_e$.
In Ref.~\cite{Forestell:2018txr} the second and higher order terms in the square bracket are neglected for $y_e<0.05$.  
However, numerically, the second term in the square bracket is $\sim -24y_e$, and is sizable when $y_e\sim O(10^{-2})$. 
With neglecting higher order terms in $y_e$,
the energy loss rate of the electron is overestimated, which may
affect the photon spectrum induced by the decay of $X$.

Once the photon spectrum is obtained, the effects of the
photodissociation processes can be included into the Boltzmann
equations governing the evolution of light elements.  The effects of
the photodissociation are taken into account by adding the following
terms into the scattering terms in the Boltzmann equations:
\begin{align}
  \left[ \frac{d n_{A_i}}{dt} \right]_{\rm photodiss}
  \equiv
  - n_{A_i} \sum_j \Gamma_{A_i\rightarrow A_j}^{\rm (photodiss)}
  + 
  \sum_j n_{A_j} \Gamma_{A_j\rightarrow A_i}^{\rm (photodiss)},
\end{align}
where $n_{A_i}$ is the number density of the nuclear species $A_i$,
and
\begin{align}
  \Gamma_{A_j\rightarrow A_i}^{\rm (photodiss)} \equiv
  \int d E_\gamma
  \sigma_{A_i \rightarrow A_j} (E_\gamma)
  f_{\gamma} (E_\gamma),
\end{align}
with $\sigma_{A_i \rightarrow A_j}$ being the cross section for the
process $\gamma+A_i\rightarrow A_j+\cdots$.  The photodissociation
processes included in our analysis are summarized in
Table~\ref{table:photodis}.  In our code, the dissociation processes
of $^6$Li, $^7$Li, and $^7$Be are included.  Because the abundances of
these elements are much smaller than those of lighter ones, they do
not affect the discussion in Section \ref{sec:results} to derive the
upper bounds on the primordial abundance of $X$.  In discussing the
implication to the $^7$Li problem in Section \ref{sec:liprob}, on the
other hand, these processes (in particular, the dissociation of
$^7$Be) play an important role.

\begin{table}[t]
    \begin{center}
        \begin{tabular}{lll} 
            \hline\hline
            Reaction & Error& Reference \\ 
            \hline
            $\gamma + {\rm D}\rightarrow n + p$
            & 6\ \%
            & \cite{Evans} \\
            $\gamma + {\rm T} \rightarrow n + {\rm D}$
            & 14\%
            & \cite{ZP208-129,PRL44-129}\\
            $\gamma + {\rm T} \rightarrow  p + n + n$
            &  7\% 
            & \cite{PRL44-129}\\
            $\gamma + {\rm ^3He} \rightarrow p + {\rm D}$
            & 10\% 
            & \cite{PL11-137}\\
            $\gamma + {\rm ^3He} \rightarrow p + p + n$
            & 15\%
            & \cite{PL11-137}\\
            $\gamma + {\rm ^4He} \rightarrow p + {\rm T}$
            &  4\%
            &\cite{SJNP19-589} \\
            $\gamma + {\rm ^4He} \rightarrow n + {^3{\rm He}}$
            &  5\%
            & \cite{CJP53-802,PLB47-433} \\
            $\gamma + {\rm ^4He} \rightarrow p + n + {\rm D}$
            & 14\%
            & \cite{SJNP19-589} \\ 
            $\gamma + {\rm ^6Li} \rightarrow {\rm anything}$
            &  4\%
            & \cite{SJNP5-349} \\ 
            $\gamma + {\rm ^7Li} \rightarrow n + {^6{\rm Li}}$
            &  4\%
            & \cite{Berman} \\
            $\gamma + {\rm ^7Li} \rightarrow {\rm anything}$
            &  9\%
            & \cite{SJNP5-344} \\ 
            $\gamma + {\rm ^7Be} \rightarrow {^3{\rm He}}+{^4{\rm He}}$
            &  9\%
            & \cite{Ishida:2014wqa} \\
            $\gamma + {\rm ^7Be} \rightarrow p + {^6{\rm Li}}$
            &  4\%
            & \cite{Cyburt:2002uv}  \\
            $\gamma + {\rm ^7Be} \rightarrow p + p + n +{^4{\rm He}}$
            &  9\%
            & \cite{Cyburt:2002uv}  \\
            \hline\hline
        \end{tabular}
        \caption{Photodissociation processes included in our analysis.
          The errors of each cross sections adopted in our Monte Carlo
          analysis are also shown.  }
        \label{table:photodis}
    \end{center}
\end{table}

%%%%%%%%%%%%%%%%%%%%%%%%%%%%%%%%%%%%%%%%%%%%%%%%%%%%%%%%%%%%%%%%%%%%%%
\section{Late-time entropy production due to the decay of \texorpdfstring{$X$}{}}
\label{sec:entropyprod}
%%%%%%%%%%%%%%%%%%%%%%%%%%%%%%%%%%%%%%%%%%%%%%%%%%%%%%%%%%%%%%%%%%%%%%

In this section, we discuss the so-called ``dilution factor'' of the
relic particle by the late-time entropy production due to massive
particles decaying into 100 $\%$ of electromagnetic energy, e.g.,
photons or electrons without producing neutrinos. In this case, we can
approximately express the dilution factor to be
\begin{eqnarray}
    \label{eq:dilutin_factor}
    \Delta_{\rm dilution} &=& 1 + \frac{\Delta s}{s}  
    = \left(\frac{T_{\gamma a}}{T_{\gamma b}} \right)^{3}
    = \left( 1 + \frac{\Delta \rho_{\gamma}}{\rho_{\gamma}} \right)^{3/4},
       \nonumber
\end{eqnarray}
where $T_{\gamma a}$ and $T_{\gamma b}$ mean approximately the photon
temperature just after and just before the entropy production,
respectively. Here we assume that electromagnetic particles are
immediately thermalized, which is approximately validated well before
the recombination epoch. In this expression, $\Delta s$ is the
increase of the entropy density $s$ due to the decaying massive
particles. Then the baryon to photon ratio is diluted after the
entropy production to be
\begin{eqnarray}
  \label{eq:etaAB}
  \eta(T_{\gamma a}) = \frac{\eta(T_{\gamma b}) }{\Delta_{\rm dilution} },
\end{eqnarray}
where $\eta(T_{\gamma a})$ could be observed by cosmic microwave
background (CMB) experiments such as Planck
(See~Eq.~(\ref{eq:etaobs})). Actually we set an initial value of
$\eta$ to be $\eta(T_{\gamma b})$ well in advance before the entropy
production, e.g., at $T \gg 10$~MeV with simultaneously fitting
$\eta(T_{\gamma a})$ to be the required value in
Eq.~(\ref{eq:etaobs}). In Appendix~\ref{sec:analyticD}, we show an
analytical calculation of $\Delta_{\rm dilution}$ in a general setup
in the parameter spaces.  The analytical formula are consistent with
actual computations of $\Delta_{\rm dilution}$ within a few percent
accuracy in the parameter space where we are considering in the
current study, e.g., for $\Delta_{\rm dilution} \lesssim {\cal
  O}(10)$.

After $T \ll 1$~MeV, if the copious entropy is produced by injection
of electromagnetic energy due to decaying massive particles, the cosmic
neutrinos cannot be thermalized.\footnote{In MeV-scale reheating
  temperature scenarios it is known that the effective number of
  neutrino species for active three-flavor neutrinos should have become
  much smaller than three, e.g., concretely $N_{\nu} < 0.1$ for the
  reheating temperature   $T_R <   0.5$~MeV.~\cite{Kawasaki:1999na,Kawasaki:2000en,Hannestad:2004px,Ichikawa:2005vw,deSalas:2015glj,Hasegawa:2019jsa}
  (See Fig.2 of Ref.\cite{Kawasaki:1999na}). }

Due to such a sizable dilution factor for $t \gg 1$~sec, the effective
number of neutrino species is also modified to be
$3.046 \to 3.046 - |\Delta N_{\nu}|$, which is
constrained by the CMB observations (\ref{eq:NeffnuObs}) as
\begin{eqnarray}
  \label{eq:DeltaNnu}
    3.046 - |\Delta N_{\nu}| = 3.046 \Delta_{\rm dilution}^{-4/3} > N_{\nu,{\rm low}}^{\rm eff},
\end{eqnarray}
where $N_{\nu,{\rm low}}^{\rm eff} = 2.55$ at 95 $\%$ C.L. as is shown
in (\ref{eq:NeffnuObs}). Thus, we obtain an upper bound on the
modification of $N_{\nu}^{\rm eff}$,
\begin{eqnarray}
  \label{eq:DeltaNnuObs}
    |\Delta N_{\nu}| < 3.046 - N_{\nu,{\rm low}}^{\rm eff} \simeq 0.50,
\end{eqnarray}
which leads to an upper bound on the dilution factor
\begin{eqnarray}
  \label{eq:DeltaDilutionObs}
    \Delta_{\rm dilution} < \left( \frac{3.046}{N_{\nu,{\rm low}}^{\rm eff}} \right)^{3/4} \simeq 1.14,
\end{eqnarray}
at 95 $\%$ C.L. 
By using the constraint on $\Delta N_{\nu}$ shown in~(\ref{eq:DeltaNnuObs}), we can
constrain parameters of the lifetime and the abundance of the decaying
massive particles.

%%%%%%%%%%%%%%%%%%%%%%%%%%%%%%%%%%%%%%%%%%%%%%%%%%%%%%%%%%%%%%%%%%%%%%
\section{Results}
\label{sec:results}
%%%%%%%%%%%%%%%%%%%%%%%%%%%%%%%%%%%%%%%%%%%%%%%%%%%%%%%%%%%%%%%%%%%%%%

Now we show BBN constraints on the injections of sub-GeV photons and
electrons.  In order to calculate abundances of light elements with
their theoretical errors, we execute the Monte Carlo estimation by
including errors in $\eta$, lifetime of neutron, and reaction rates in
both the standard processes in SBBN and the non-standard
photodissociation processes.

\subsection{Injections of a high-energy line photon}

\begin{figure}
\begin{center}
\includegraphics[scale=0.36]{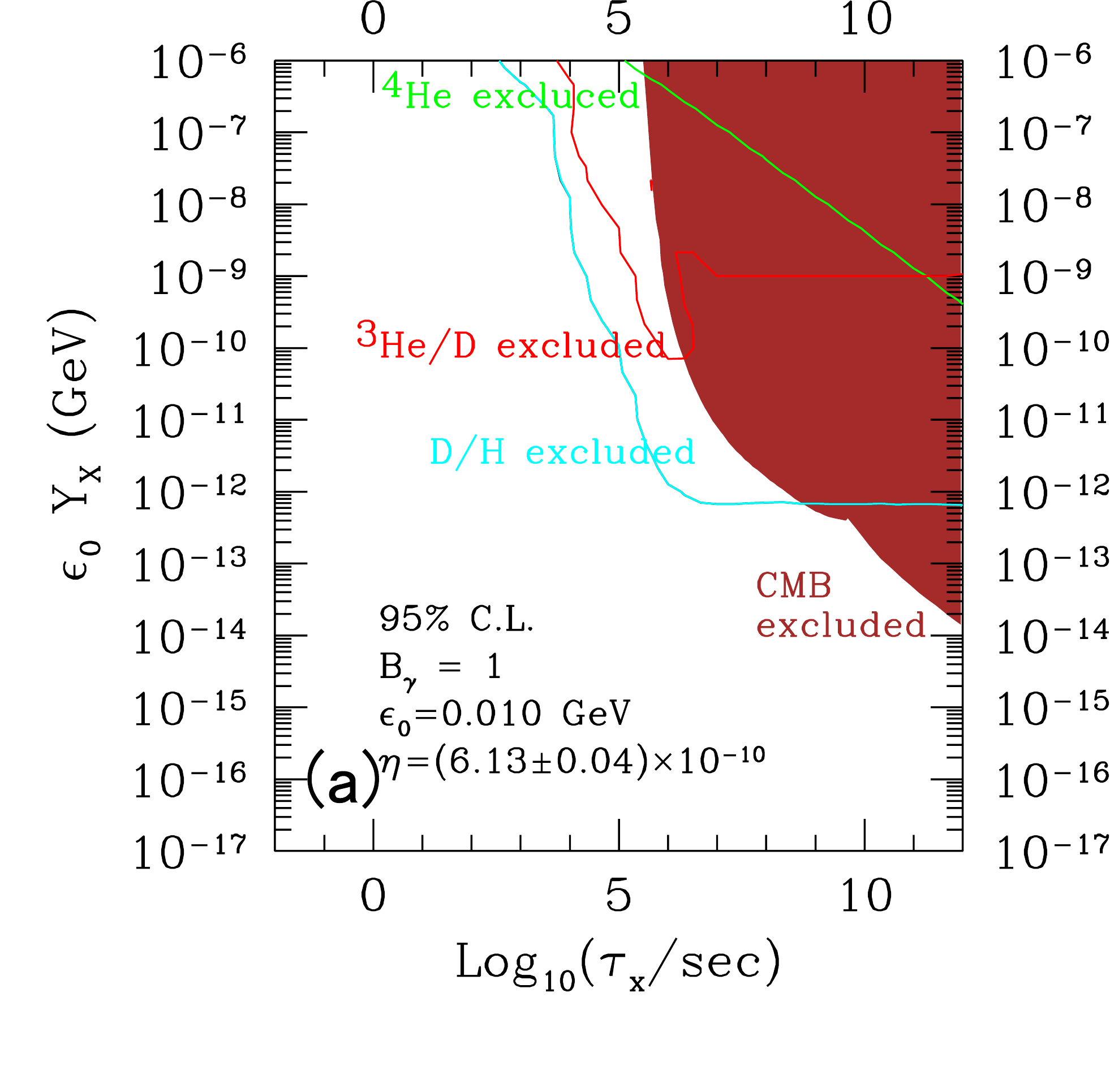}
\includegraphics[scale=0.36]{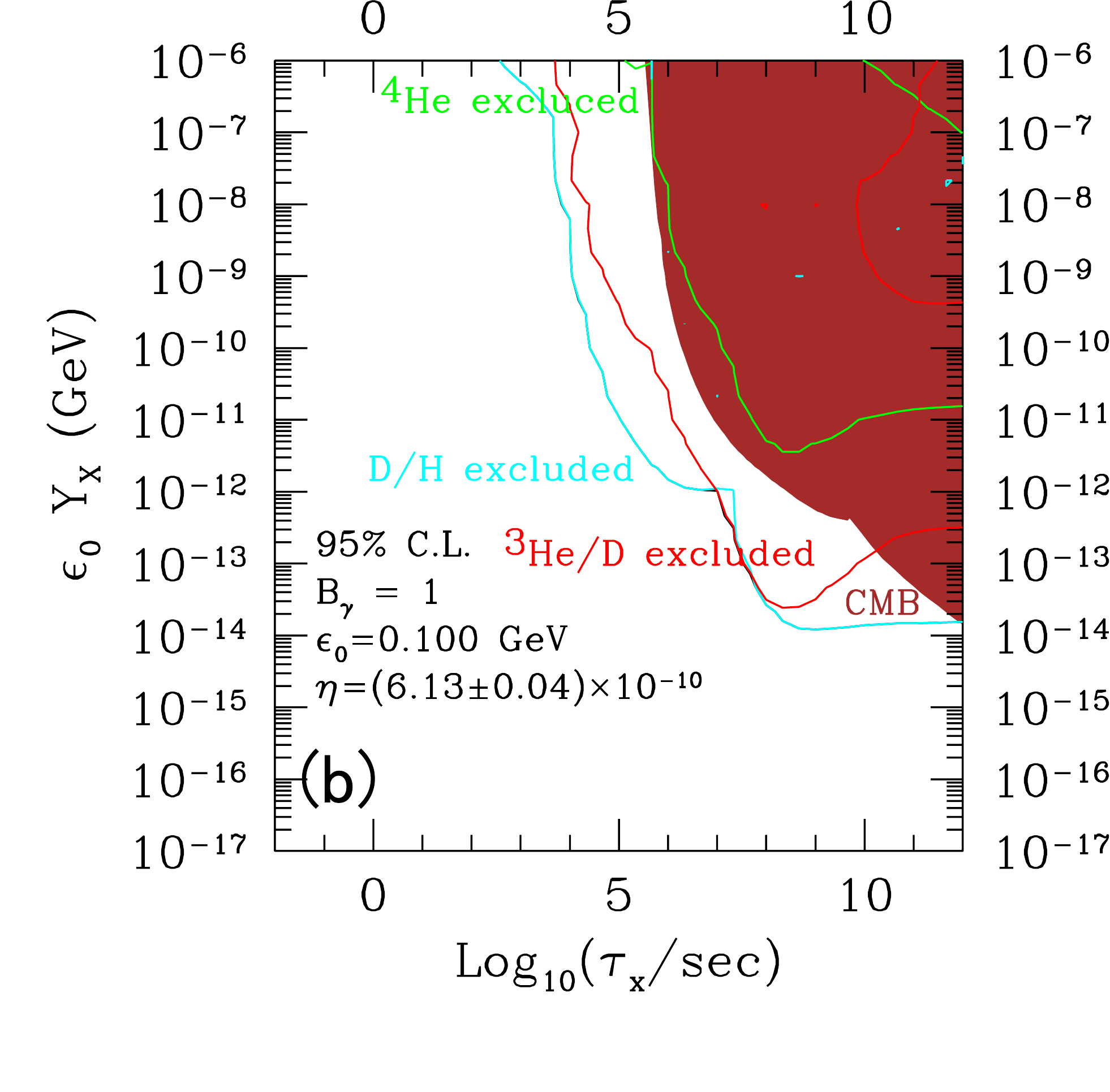}
\includegraphics[scale=0.36]{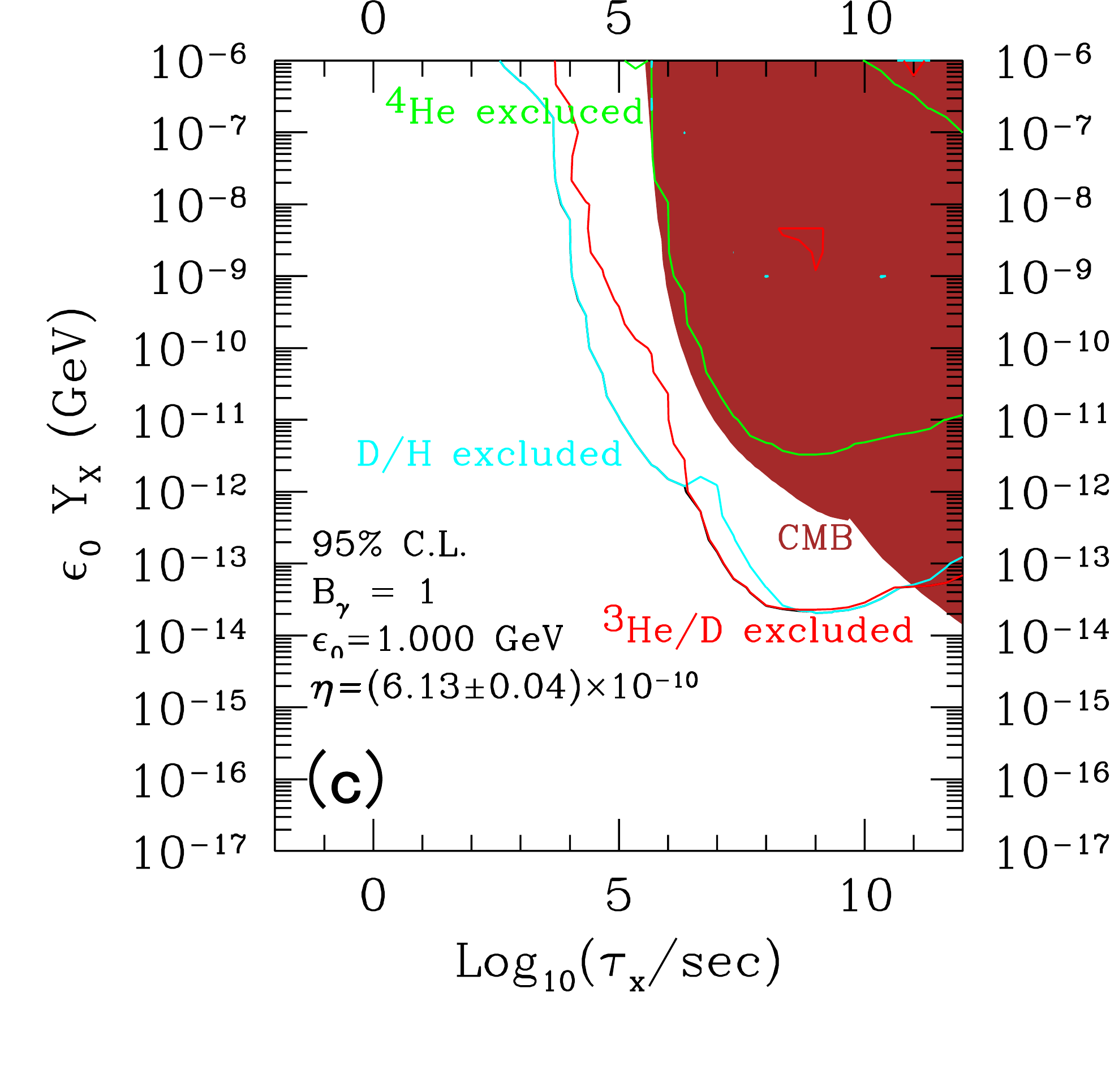}
\includegraphics[scale=0.36]{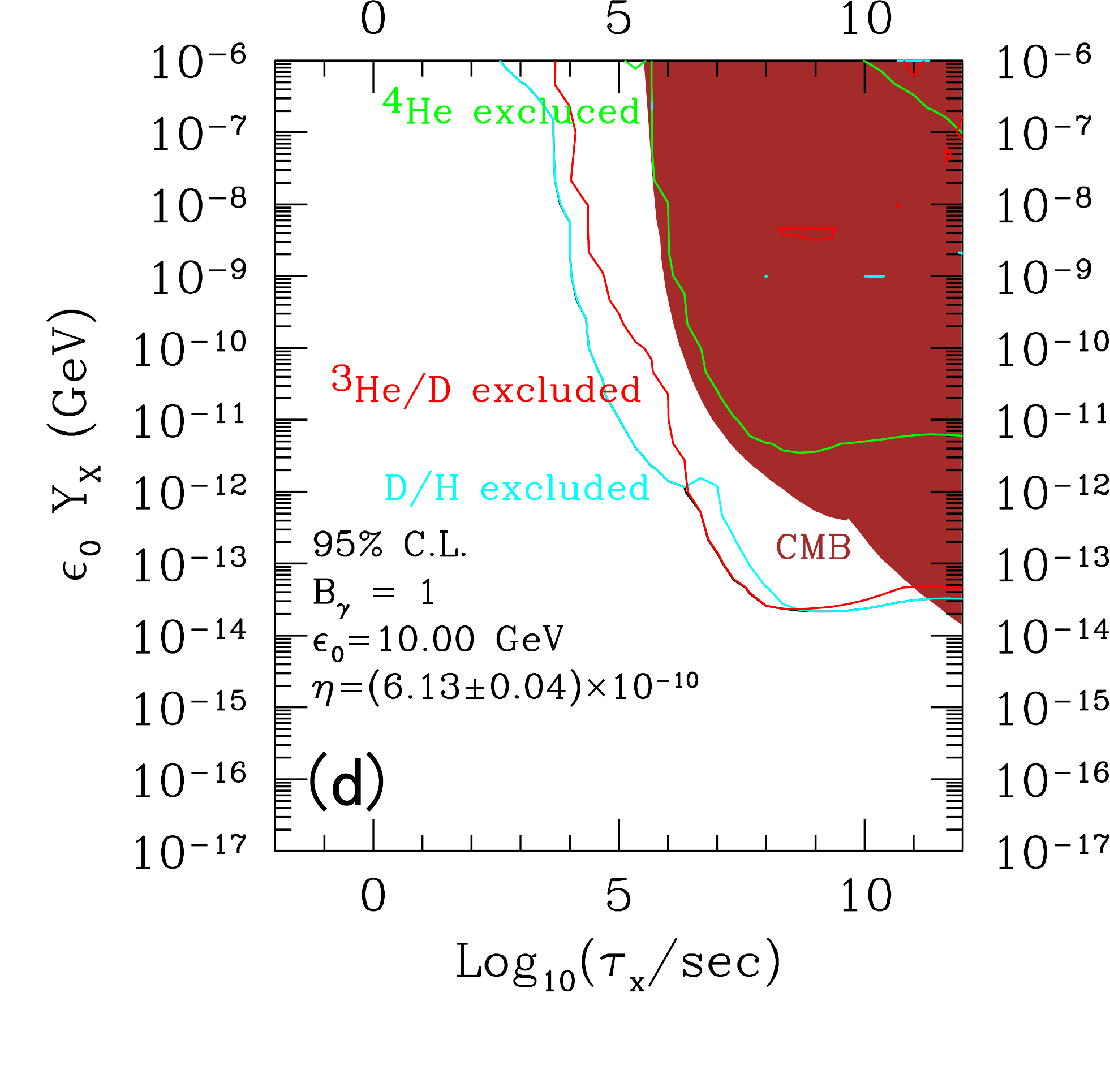}
\vspace{-1.0cm}
\caption{Upper bounds on $\epsilon_0 Y_X$ at 95$\%$ C.L. as a function
  of lifetime for the energy of a high-energy injected photon, (a)
  $\epsilon_0 = 10^{-2}$~GeV, (b) $10^{-1}$~GeV, (c) 1~GeV and (d)
  $10\ {\rm GeV}$. We executed Monte Carlo runs. The lines denote the
  bounds from $^3$He/D (red), Y$_p$ (green), D/H (cyan), and the
  $\mu$- and y-distortion of CMB (brown), respectively.}
\label{fig:myxG}
\end{center}
\end{figure}

First, let us consider the case of sub-GeV line photon.  In
Fig.\ \ref{fig:myxG}, we show the upper bounds on $\epsilon_0Y_X$ as
functions of the lifetime of the unstable particle.  For comparison,
we also show constraints for the case of $\epsilon_0=10\ {\rm GeV}$.
Each line shows constraint from D (cyan), $^3{\rm He}/{\rm D}$ (red),
or $Y_p$ (green).  In the figure, for comparison, we also show the
constraint from the CMB distortion, adopting the result given
in~\cite{Dimastrogiovanni:2015wvk}.\footnote
{For earlier works on the
constraint from the CMB distortion, see \cite{Hu:1993gc,
  Chluba:2011hw, Chluba:2013wsa, Fixsen:1996nj, Chluba:2013pya}.}

For the lifetime shorter than $\sim 10^6\ {\rm sec}$, D imposes the
most stringent constraint on the primordial abundance of $X$.  For
such a short lifetime, the threshold for the photon-photon pair
creation (i.e., $\gamma+\gamma_{\rm BG}\rightarrow e^+e^-$), which is
$\sim m_e^2/22T$ with $m_e$ being the electron mass
\cite{Kawasaki:1994sc}, is smaller than the thresholds of the
dissociation processes of $^4{\rm He}$.  Thus, the constraint is
mainly due to the photodissociation of D.  For longer lifetime, the
photodissociation of background $^4{\rm He}$ may result in the
overproductions of D and $^3{\rm He}$.  When the energy of the
injected photon is high enough (i.e., $\epsilon_0\gtrsim 1\ {\rm
  GeV}$), the constraints from D and $^3{\rm He}/{\rm D}$ are
comparable for high enough injection energy ($\epsilon_0\gtrsim
O(100\ {\rm MeV})$).  For smaller injection energy
($20~\text{MeV}\lesssim \epsilon_0 \lesssim O(100)$~MeV), the $^3{\rm
  He}/{\rm H}$ constraint becomes weaker so that the D constraint
dominates the BBN bound on the primordial abundance of $X$.
Furthermore, when the injection energy is smaller than $\sim 20\ {\rm
  MeV}$ (i.e.,the threshold energy for the $^4{\rm He}$ dissociation),
the D constraint becomes significantly weakened because, in this case,
the photodissociation of $^4{\rm He}$ cannot occur and the
overproduction of D due to the $^4{\rm He}$ dissociation becomes
irrelevant.  We also comment here that, for the injection energy below
the threshold of the $^4{\rm He}$ dissociation, the constraint from
the CMB distortion gives a stronger constraint than the BBN for
$\tau_X \gtrsim 10^9$~sec.

%%%%%%%%%%%%%%%%%%%%%%%%%%%%%%%%%%%%%%%%%%%%%%%%%%%%%%%%%%%%%%%%%%%%%%
%%%%%%%%%%%%%%%%%%%%%%%%%%%%%%%%%%%%%%%%%%%%%%%%%%%%%%%%%%%%%%%%%%%%%%

\subsection{Injections of a high-energy line electron}

%%%%%%%%%%%%%%%%%%%%%%%%%%%%%%%%%%%%%%%%%%%%%%%%%%%%%%%%%%%%%%%%%%%%%%

\begin{figure}
\begin{center}
\includegraphics[scale=0.36]{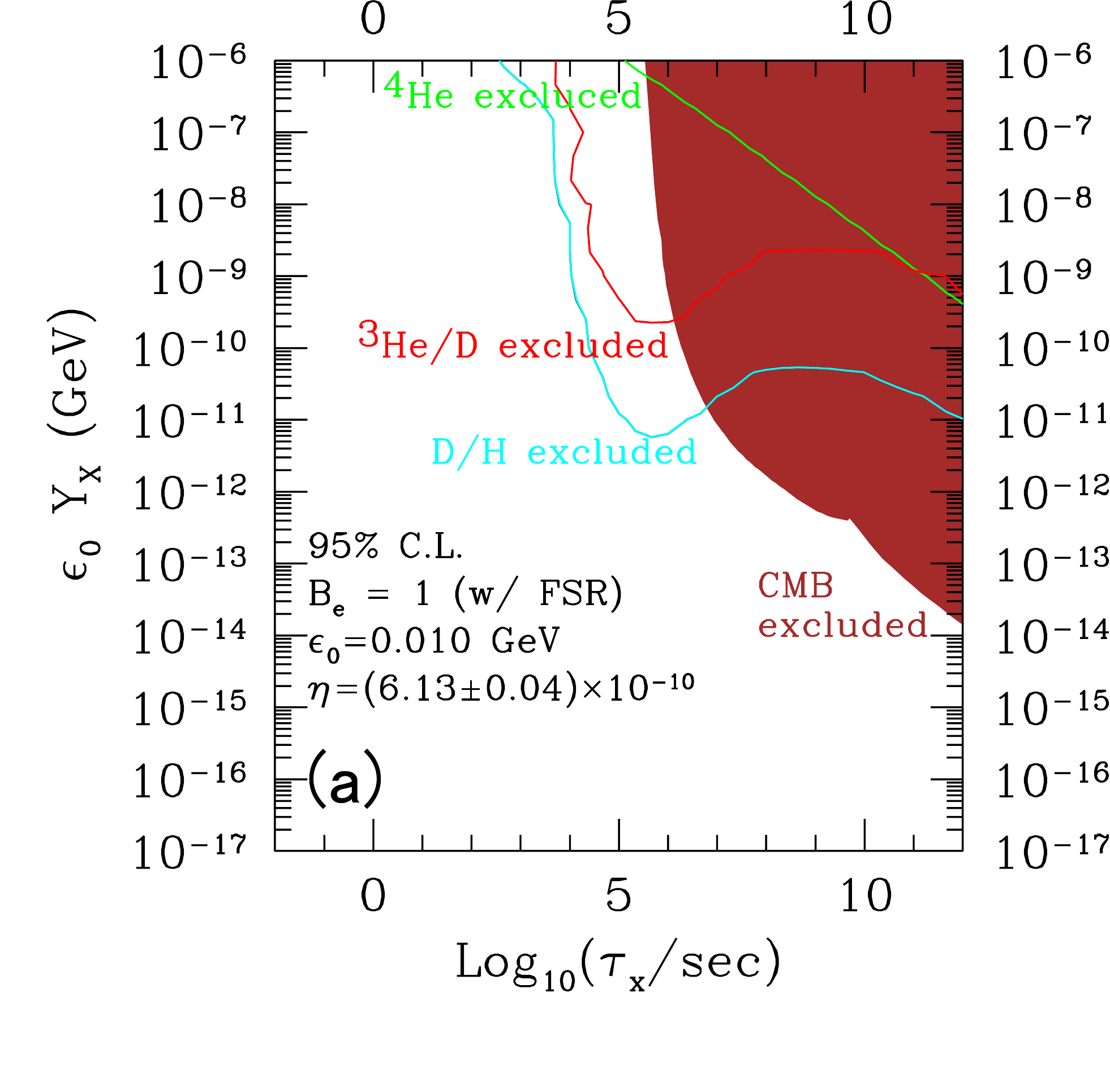}
\includegraphics[scale=0.36]{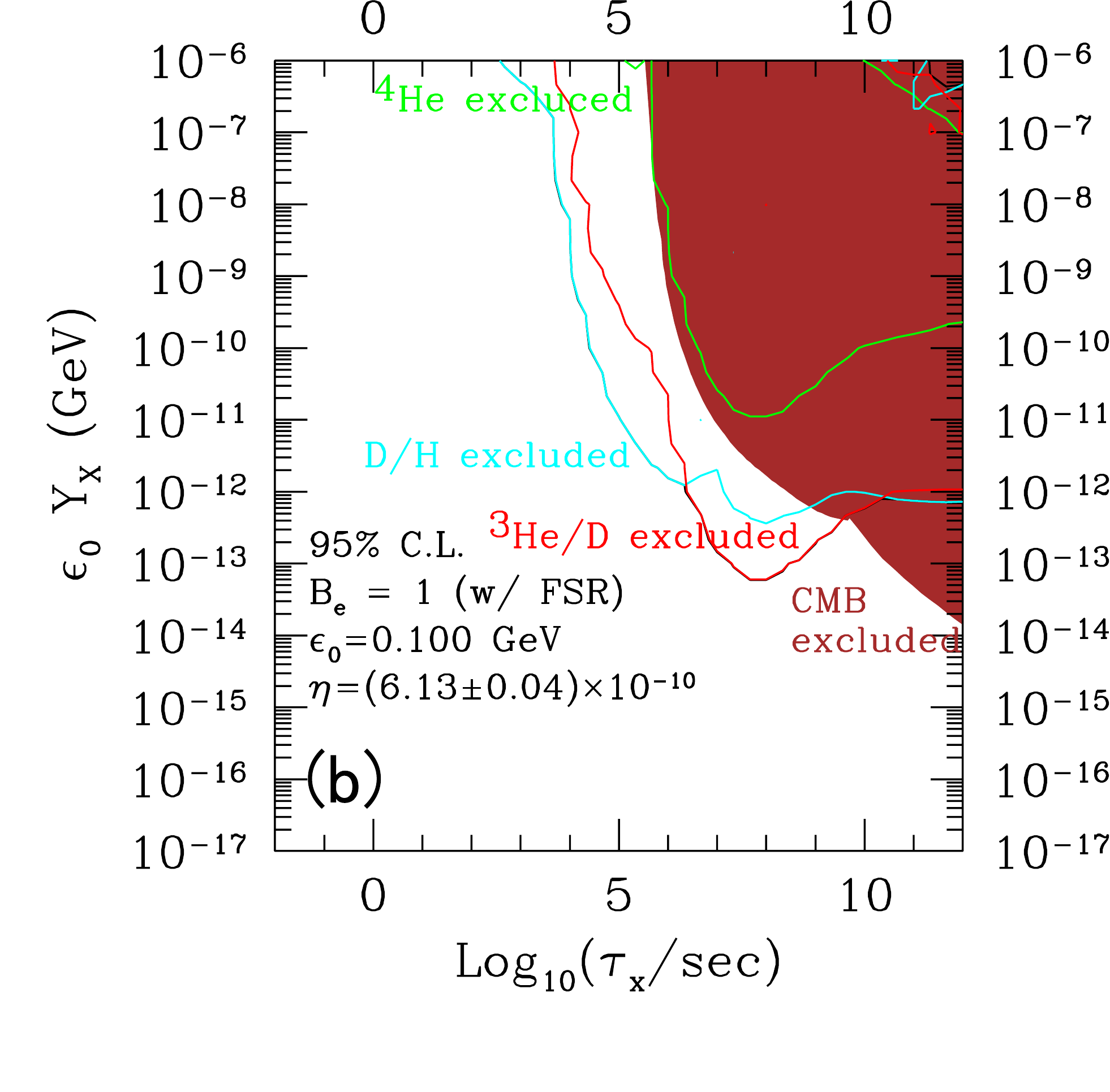}
\includegraphics[scale=0.36]{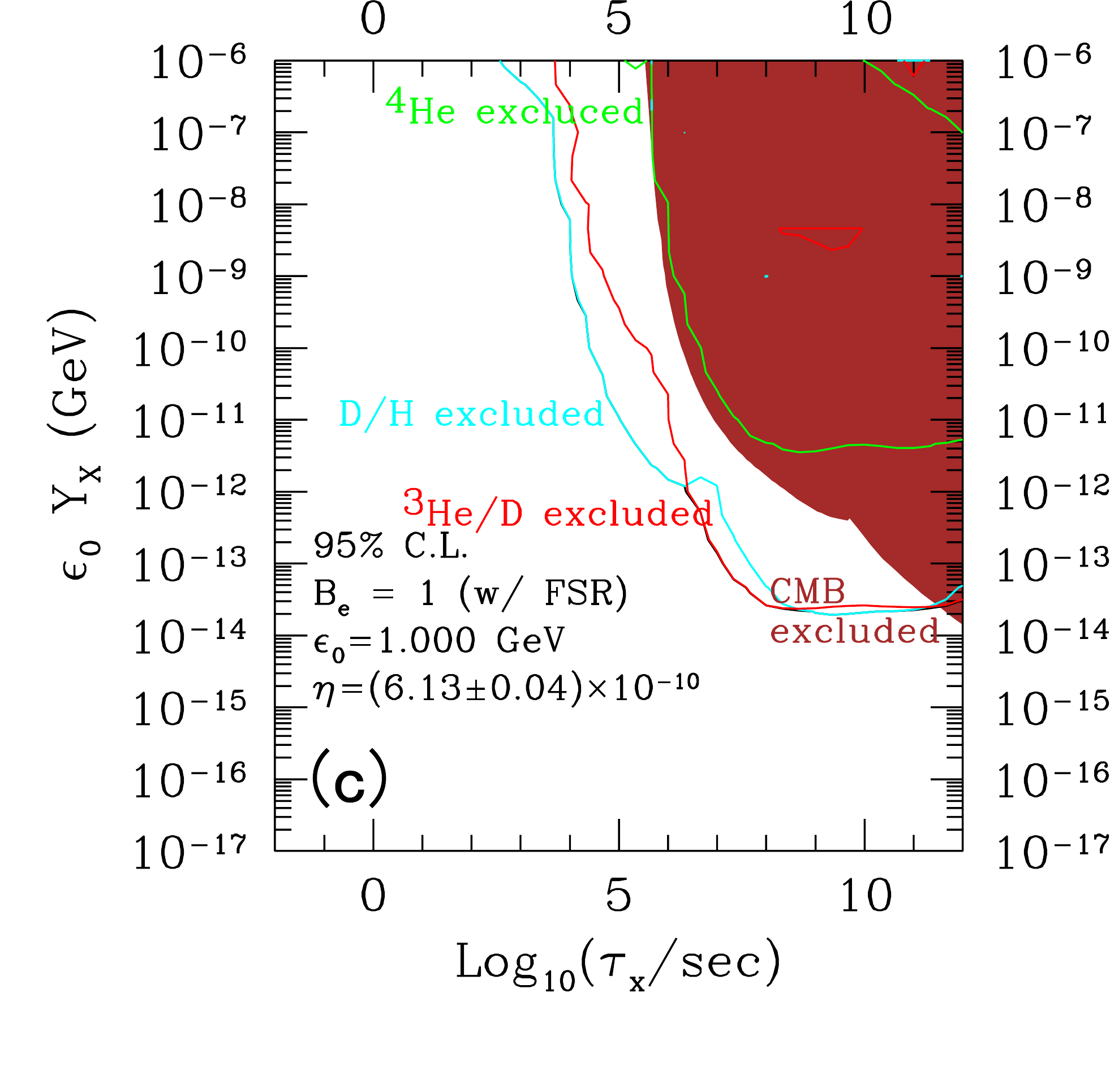}
\includegraphics[scale=0.36]{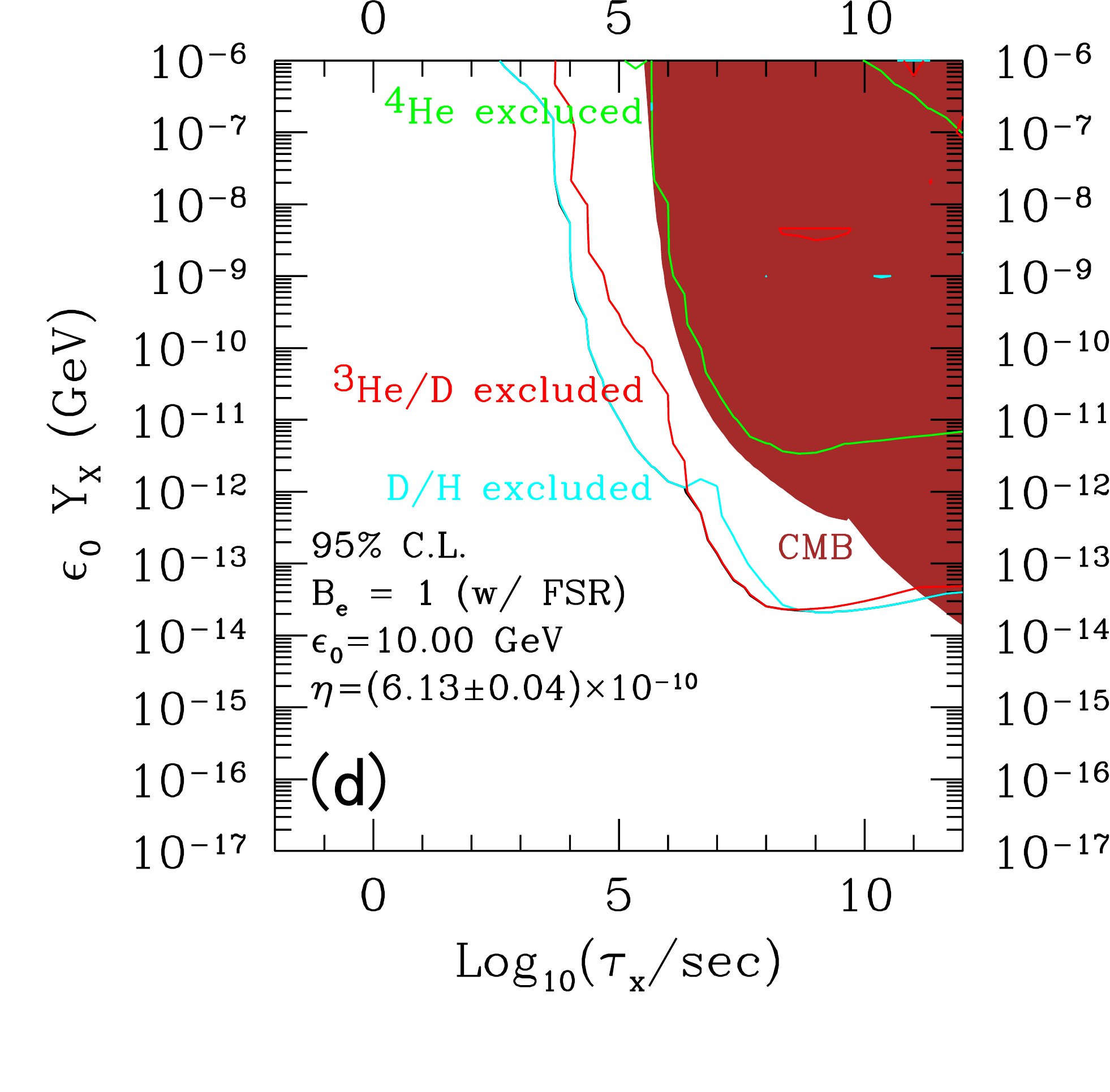}
\vspace{-1.0cm}
\caption{Upper bounds on $\epsilon_0 Y_X$ at 95$\%$ C.L. as a function
  of lifetime $\tau_X$ for the energy of a high-energy injected
  electron, (a) $\epsilon_0 = 10^{-2}$~GeV, (b) $10^{-1}$~GeV, (c)
  1~GeV and (d) $10\ {\rm GeV}$.  Effects of the FSR are included
  in the calculation of the photon and electron spectrum.  We executed
  Monte Carlo runs. The lines denote the bounds from $^3$He/D (red),
  Y$_p$ (green), D/H (cyan), and the $\mu$- and y-distortion of CMB
  (brown), respectively.}
\label{fig:myxEFSR}
\end{center}
\end{figure}

Next, we study the case of a high-energy electron injection.  In
Fig.\ \ref{fig:myxEFSR}, we show the upper bounds on $\epsilon_0Y_X$
as functions of the lifetime of the unstable particle.  As we can see,
when the injection energy is high enough (i.e, $\epsilon_0\gtrsim
O(100)\ {\rm MeV}$), the upper bounds on $\epsilon_0Y_X$ are almost
unchanged from those for the case of the high energy photon injection;
for such case, the photon spectrum is mostly determined by the total
amount of the injected energy in the form of electromagnetic
particles, and hence the bounds are insensitive to the species of the
injected particles (i.e., $\gamma$ or $e^\pm$).  For a smaller
injection energy, on the contrary, the constraint becomes weaker
compared to the case of the high-energy photon injection in particular
when the lifetime is relatively long.  This is due to the suppression
of the photon spectrum for the case where the electromagnetic shower
is mainly initiated by the FSR photons (see Section \ref{sec:shower}).
We also note here that the CMB constraint becomes stronger than the
BBN constraint for a long lifetime.  In particular, for the injection
energy lower than the threshold of the $^4$He dissociation, the CMB
distortion gives more stringent constraint for $\tau_X\gtrsim
10^7\ {\rm sec}$.

\begin{figure}
\begin{center}
\includegraphics[scale=0.36]{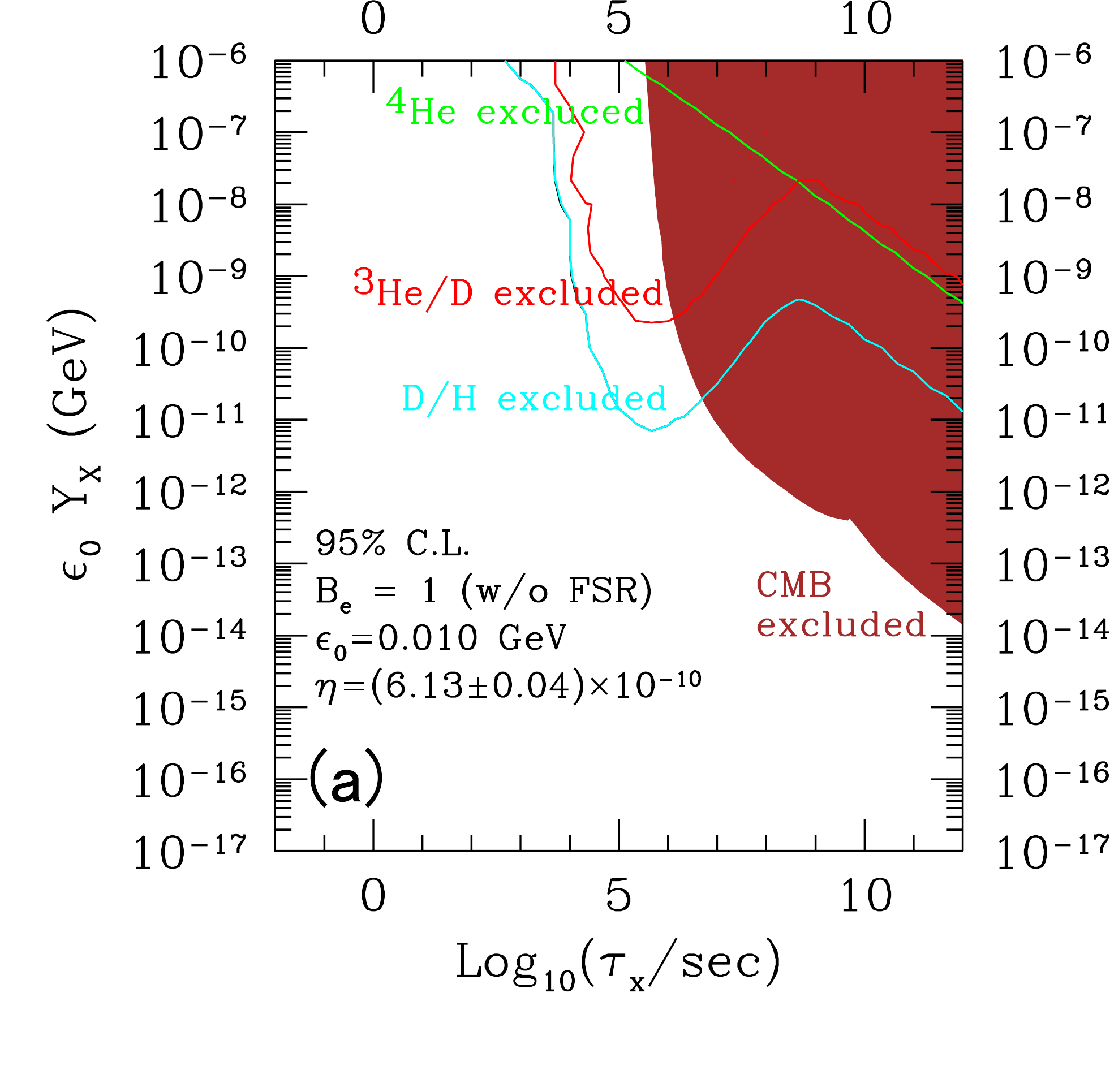}
\includegraphics[scale=0.36]{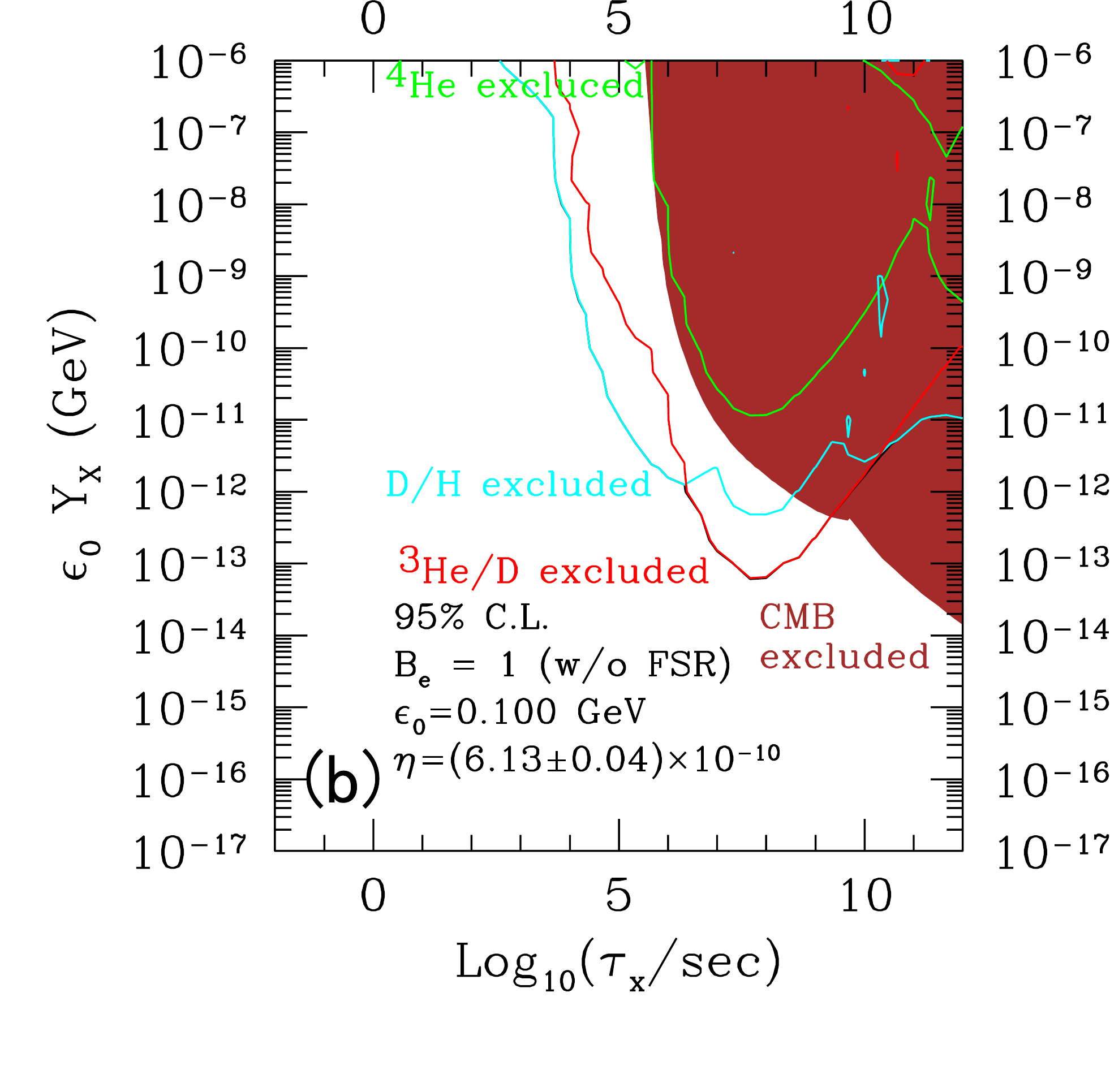}
\vspace{-1.0cm}
\caption{Upper bounds on $\epsilon_0 Y_X$ at 95$\%$ C.L. as a function
  of lifetime $\tau_X$ for the energy of a high-energy injected
  electron, (a) $\epsilon_0 = 10^{-2}$ and (b) $10^{-1}~{\rm GeV}$ omitting
  effects of the FSR associated with electrons in the final states. We
  executed Monte Carlo runs. The lines denote the bounds from $^3$He/D
  (red), Y$_p$ (green), D/H (cyan), and the $\mu$-
  and y-distortion of CMB (brown), respectively.}
\label{fig:myxEnoFSR}
\end{center}
\end{figure}

For comparison, in Fig.\ \ref{fig:myxEnoFSR}, we show how the
constraints behave if we neglect the effects of FSR, taking
$\epsilon_0 = 10$ and $100\ {\rm MeV}$.  (We have checked that the
constraints are almost unchanged for
$\epsilon_0\gtrsim 1\ {\rm GeV}$.)  We can see that the BBN constraint
is weakened for $\epsilon_0=100\ {\rm MeV}$ when the lifetime is
longer than $\sim 10^{8}\ {\rm sec}$.  For the case of
$\epsilon_0=10\ {\rm MeV}$, the BBN constraints become weaker for
$\tau_X\gtrsim 10^{6}\ {\rm sec}$.  However, for longer lifetime, the
constraints becomes almost unchanged even if we neglect the effects of
FSR.  This is due to the fact that, for such a parameter region, the
BBN constraint comes mainly from the change of the baryon-to-photon
ratio due to the dilution. As discussed in Section
\ref{sec:entropyprod}, the emission of the electromagnetic particle
due to the decaying $X$ induces entropy production which affects the
value of the baryon-to-photon ratio.  Thus, with the present value of
$\eta$ being fixed as Eq.\ \eqref{eq:etaobs}, the baryon-to-photon
ratio at the time of the BBN epoch is larger than it when the effect
of the entropy production is significant. The light element abundances
are sensitive to the value of $\eta$, and also the BBN calculation
based on the value of $\eta$ given in Eq.\ \eqref{eq:etaobs}. Because
the theoretical values of light element abundances are more or less
consistent with observed values, the parameter regions, in which large
entropy production is induced, are excluded by the observations.  The
FSR does not affect such a constraint, which is the reason why the
constraints become insensitive to the inclusion of the FSR for
$\epsilon_0=10\ {\rm MeV}$ and long-enough lifetime.  (Notice that
this is the case only when $\epsilon_0\lesssim 20\ {\rm MeV}$, i.e,
the threshold energy of $^4$He.)

%%%%%%%%%%%%%%%%%%%%%%%%%%%%%%%%%%%%%%%%%%%%%%%%%%%%%%%%%%%%%%%%%%%%%%

%%%%%%%%%%%%%%%%%%%%%%%%%%%%%%%%%%%%%%%%%%%%%%%%%%%%%%%%%%%%%%%%%%%%%%

%%%%%%%%%%%%%%%%%%%%%%%%%%%%%%%%%%%%%%%%%%%%%%%%%%%%%%%%%%%%%%%%%%%%%%
\subsection{Comparison with previous works}
%%%%%%%%%%%%%%%%%%%%%%%%%%%%%%%%%%%%%%%%%%%%%%%%%%%%%%%%%%%%%%%%%%%%%%

Here we compare our results with those reported by earlier works on
sub-GeV massive particles decaying into electromagnetic daughter
particles
\cite{Poulin:2015woa,Poulin:2015opa,Forestell:2018txr,Acharya:2019uba}.
Without adopting the universal photon spectrum~\cite{Kawasaki:1994sc},
we have to solve the Boltzmann equations. As shown in
Fig.~\ref{fig:spect0010} and Fig.~\ref{fig:spect0100}, our spectra for
both the nonthermal photons and electrons are consistent with those in
~\cite{Forestell:2018txr} approximately within a factor of two. Before
we directly compare our results with theirs for the bounds on the
light element abundances, we remark the following three points.
\begin{itemize}
\item[1.] We executed the Monte Carlo runs to evaluate theoretical
  errors of light element abundances. By performing the $\chi^2$
  analysis using both the theoretical and observational errors, we
  obtained the upper bounds on $Y_X$ as functions of $\tau_X$ at
  95$\%$ C.L.  Because of the Monte Carlo estimation of the
  theoretical uncertainties, our bounds tend to become milder by a
  factor of $\sim 2$ than those without the Monte Carlo estimation.
\item[2.] As for the observational bound on the primordial value of
  $^3$He, we adopted the upper bound on $^3$He/D~\cite{Geiss:2003}.
  On the other hand, the authors in
  Refs.~\cite{Poulin:2015woa,Poulin:2015opa,Forestell:2018txr,Acharya:2019uba}
  adopted the observational bound on $^3$He/H.  We believe an upper
  bound on the primordial value of $^3$He obtained from $^3$He/D is
  more reasonable and conservative than that obtained from
  $^3$He/H~\cite{Sigl:1995kk,Holtmann:1998gd}. That is because the
  primordial $^3$He can be destroyed in relatively small
  stars~\cite{Dearborn:1985xw,Steigman:1995ku}; it is highly uncertain
  to estimate how much $^3$He is destroyed in such stars (see also
  discussions in~\cite{Kohri:1996ke}). When we consider both
  destruction and production processes of $^3$He in stars, it is
  remarkable that the ratio $^3$He/D simply increases as a function of
  the cosmic time through chemical evolutions because D is more
  fragile than $^3$He and is destroyed whenever $^3$He is
  destroyed. The upper bound on $^3$He/D shown in
  (\ref{eq:obs_const_He3D}) gives a milder bound on $Y_X$ than that
  from $^3$He/H~$<(1.0 \pm 0.5) \times
  10^{-5}$~\cite{Forestell:2018txr,Geiss:2003}, approximately at most
  by a factor of $\sim 4$ even without executing the Monte Carlo
  estimation.
\item[3.] We took into account a dilution of baryon by the entropy
  production due to the decay of $X$. In our analysis, as an initial
  condition well before X starts to decay, we took a larger initial
  value of $\eta$ to realize the present value given in
  (\ref{eq:etaobs}). At around top-right regions in
  Figs.~\ref{fig:myxG} -- \ref{fig:myxGMeVefsr}, the light element
  abundances are calculated with the initial value of $\eta$
  significantly larger than the value given in (\ref{eq:etaobs}),
  which gives some difference between ours and the analyses without
  such modifications on $\eta$.
\end{itemize}

By considering the above three points, our constraints are consistent
with those in ~\cite{Poulin:2015woa,Poulin:2015opa,Forestell:2018txr}
approximately within a factor of two. On the other hand, compared with
the results in~Ref.~\cite{Acharya:2019uba}, it seems that deviations
are somehow much larger for the injections of the lower energy photon
($\epsilon_0 \lesssim $10 MeV).

%%%%%%%%%%%%%%%%%%%%%%%%%%%%%%%%%%%%%%%%%%%%%%%%%%%%%%%%%%%%%%%%%%%%%%
\section{Implication to the $^7$Li Problem}
\label{sec:liprob}
%%%%%%%%%%%%%%%%%%%%%%%%%%%%%%%%%%%%%%%%%%%%%%%%%%%%%%%%%%%%%%%%%%%%%%

So far, we have neglected the $^7$Li abundance in deriving the upper
bounds on the primordial abundance of the unstable particle.  This is
because, conservatively, the primordial abundance of $^7$Li is still
controversial as we have mentioned in Section \ref{sec:obs}.  If the
Spite plateau value of $^7$Li really indicates its primordial
abundance, it is highly inconsistent with the SBBN prediction, i.e.,
the $^7$Li problem.  It is notable, however, that a long-lived
particles decaying into photons may solve difficulty
\cite{Ishida:2014wqa,Poulin:2015woa}.  In this section, we assume that
the Spite plateau value \eqref{eq:Li7} corresponds to the primordial
abundance of $^7$Li and discuss how the decaying particle may solve
the $^7$Li problem.

Because the SBBN abundance of $^7$Li given in Eq.\ \eqref{eq:Li7} is
about three times larger than the observed value, the $^7$Li problem
may be solved if the dissociation processes induced by the radiatively
decaying particles reduce right amount of $^7$Li.  For the value of
baryon-to-photon ratio suggested by the Planck collaboration (see Eq.\
\eqref{eq:etaobs}), the $^7$Li in the present Universe mostly
originate from $^7$Be which decays into $^7$Li via the electron
capture in the SBBN.  Thus, if a significant amount of $^7$Be
is dissociated by energetic photons
emitted by the decay of $X$, the predicted value of the $^7$Li in the
present Universe may become consistent with the observed value.  In
order for such a solution to work, it should be also guaranteed that
the emitted photons to dissociate $^7$Be (and $^7$Li) should not cause
any harmful effects, i.e., dissociations of other light elements or
distortion of the CMB background.

Importantly, the threshold energy of the photon for the process ${\rm
  ^7Be}(\gamma,{\rm ^3He}){\rm ^4He}$ is $E^{\rm (th)}_{\rm
  ^7Be}\simeq 1.59\ {\rm MeV}$, which is lower than that of
the photodissociation of D ($E^{\rm (th)}_{\rm D}\simeq 2.22\ {\rm
  MeV}$) and $^4$He ($\sim 20\ {\rm MeV}$).  Thus, if the energy of
the injected photons is in the range of $E^{\rm (th)}_{\rm
  ^7Be}<\epsilon_0<E^{\rm (th)}_{\rm D}$, the photodissociation of
$^7$Be may occur to solve the $^7$Li problem without significantly
affecting the abundances of other light elements, as mentioned in
\cite{Ishida:2014wqa,Poulin:2015woa}.

%%%%%%%%%%%%%%%%%%%%%%%%%%%%%%%%%%%%%%%%%%%%%%%%%%%%%%%%%%%%%%%%%%%%%%
\begin{figure}[t]
\begin{center}
\includegraphics[scale=0.36]{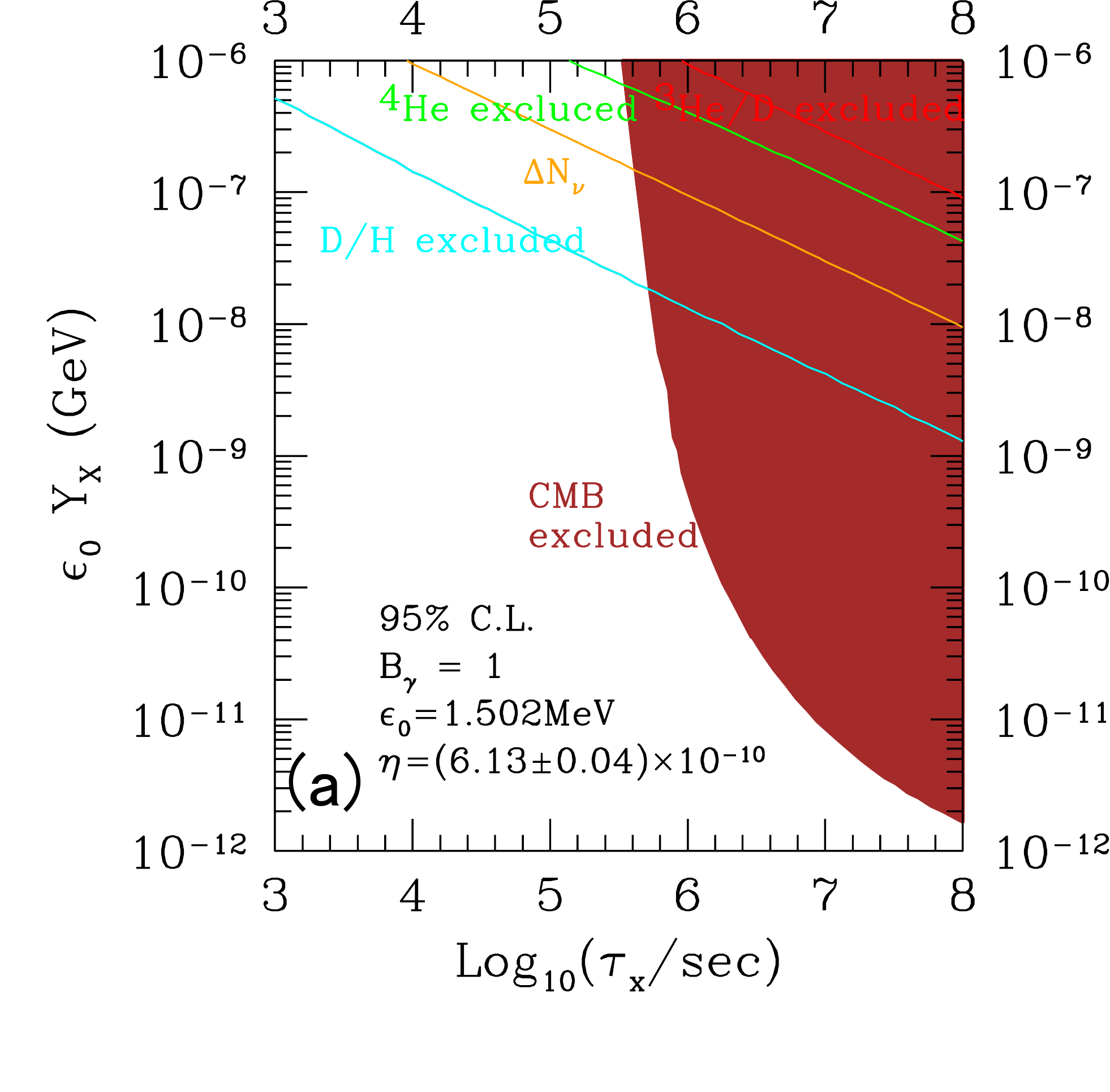}
\includegraphics[scale=0.36]{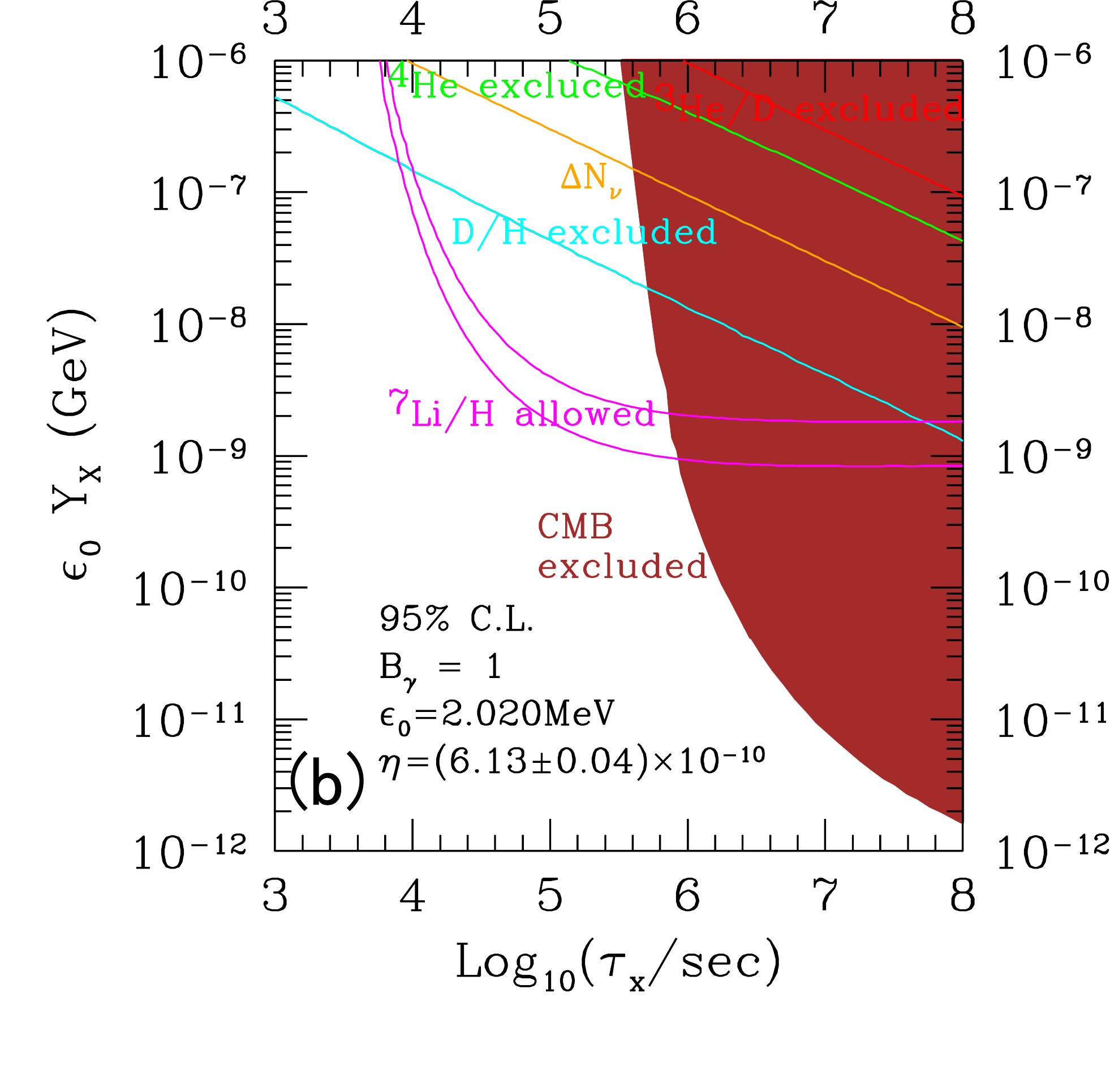}
\includegraphics[scale=0.36]{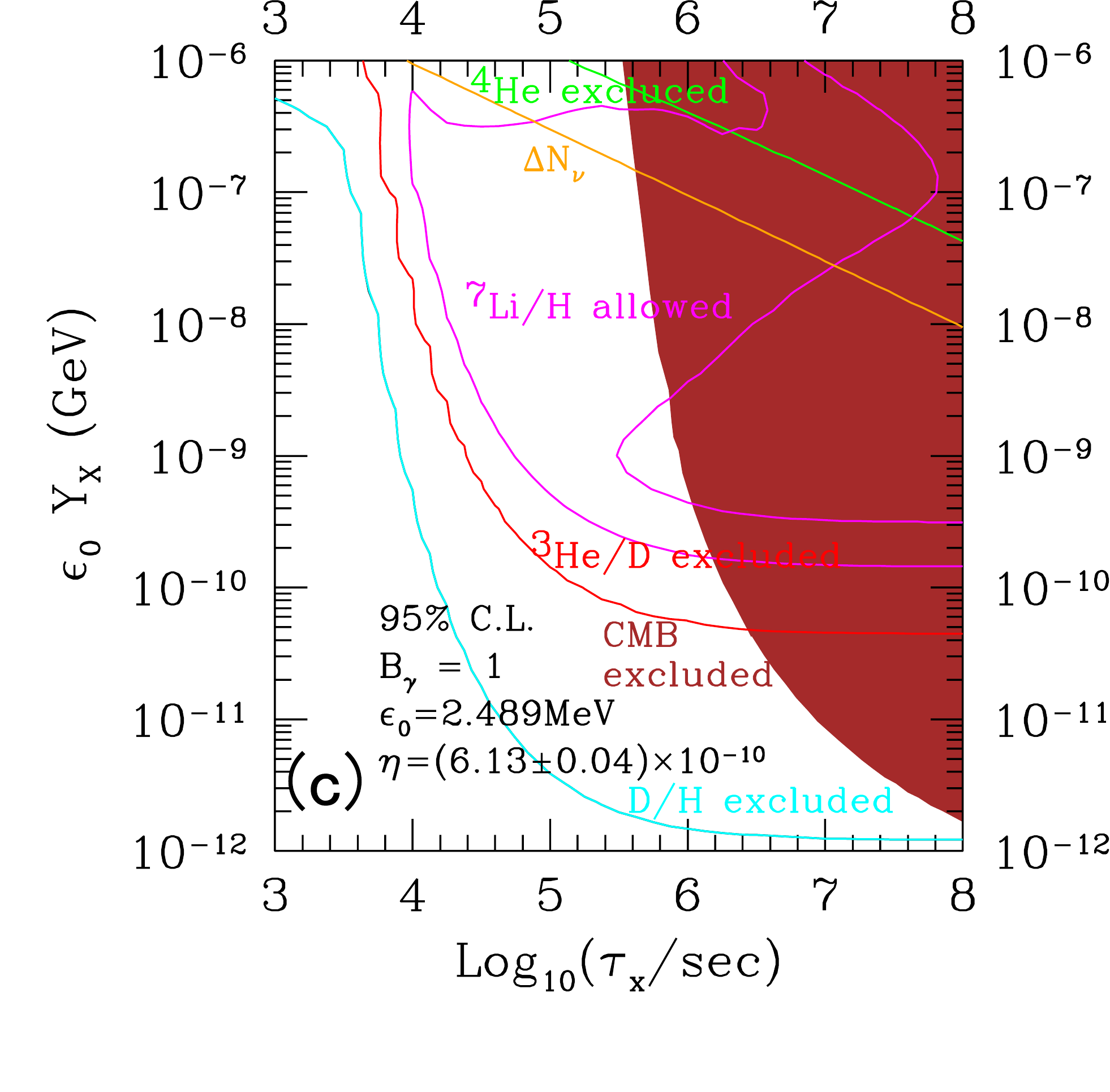}
%\vspace{-1.0cm}
\caption{Allowed regions on $\epsilon_0 Y_X$ as a function of lifetime
  ($\tau_X$) to fit $^7$Li/H (magenta) for the energy of a high-energy
  injected photon, (a) $\epsilon_0 =1.5~{\rm MeV}$, (b)
  $2.0~{\rm MeV}$, and (c) $2.5~{\rm MeV}$.  We executed Monte Carlo
  runs.The lines denote the bounds from $^3$He/D (red), Y$_p$ (green),
  D/H (cyan), and the $\mu$- and y-distortion of CMB (brown),
  respectively.}
\label{fig:myxGMeV}
\end{center}
\end{figure}
%%%%%%%%%%%%%%%%%%%%%%%%%%%%%%%%%%%%%%%%%%%%%%%%%%%%%%%%%%%%%%%%%%%%%%

To see if this scenario really works, we calculate the light element
abundances for the case of monochromatic photon injection with
$\epsilon_0=1.5$, $2.0$, and $2.5\ {\rm MeV}$.  The results are shown
in Fig.\ \ref{fig:myxGMeV}.  For $\epsilon_0=1.5\ {\rm MeV}$, the
energy of the injected photon is lower than the thresholds of all the
photodissociation processes.  Then, the constraints are from the
entropy production due to the $X$ decay which causes the change of the
baryon-to-photon ratio as well as the CMB distortion.  Taking
$\epsilon_0=2\ {\rm MeV}$, on the contrary, the photodissociation of
$^7$Be can occur.  In this case, we can see that there shows up a
parameter region which gives the $^7$Li abundance consistent with
Eq.~\eqref{eq:Li7} without too much affecting the abundances of other
light elements.  For $\tau_X\gtrsim 2\times 10^6\ {\rm sec}$, the
constraint from the CMB distortion excludes the parameter region of
our interest.  The CMB constraint is, however irrelevant for shorter
lifetime.  Consequent, for
$8\times 10^3\lesssim\tau_X\lesssim 2\times 10^6\ {\rm sec}$, we can
find a parameter region giving the $^7$Li abundance consistent with
Eq.~\eqref{eq:Li7} without conflicting the other constraints.  Here,
we remark that the photodissociation process in this case is mainly
induced by the photons just after the injection, i.e., photons with
the energy of $\epsilon_0$.  Photons after experiencing the
electromagnetic shower processes, i.e., cannot effectively induce the
photodissociation process when $\epsilon_0$ is close to the threshold
energy.  With the photon energy larger than $E^{\rm (th)}_{\rm D}$,
the photodissociation of D occurs as well as the that of $^7$Be.  A
few \% change of the D abundance due to the photodissociation results
in a disagreement with the observed value because the observational
value of the D abundance is so precise, while about $60-70\ \%$ of
$^7$Be should be photodissociated to solve the $^7$Li problem.
Because the photodissociation cross sections for these processes are
of the same order of magnitude, the D constraint excludes the
parameter region which gives right $^7$Be abundance, as indicated in
the figure.

For comparison, we also calculate the light element abundances for the
case of a monochromatic $e^\pm$ injection.  For the case of the
electron injection with $\epsilon_0\sim O(1)\ {\rm MeV}$, the
photodissociations are mostly induced by photons emitted by the FSR.
Such photon flux is not monochromatic, and is suppressed compared to
that in the case of monochromatic photon injection.  Consequently, in
the case of monochromatic $e^\pm$ injection, the $^7$Li problem is
hardly solved without conflicting the other constraints.

%%%%%%%%%%%%%%%%%%%%%%%%%%%%%%%%%%%%%%%%%%%%%%%%%%%%%%%%%%%%%%%%%%%%%%
\begin{figure}[t]
\begin{center}
\includegraphics[scale=0.36]{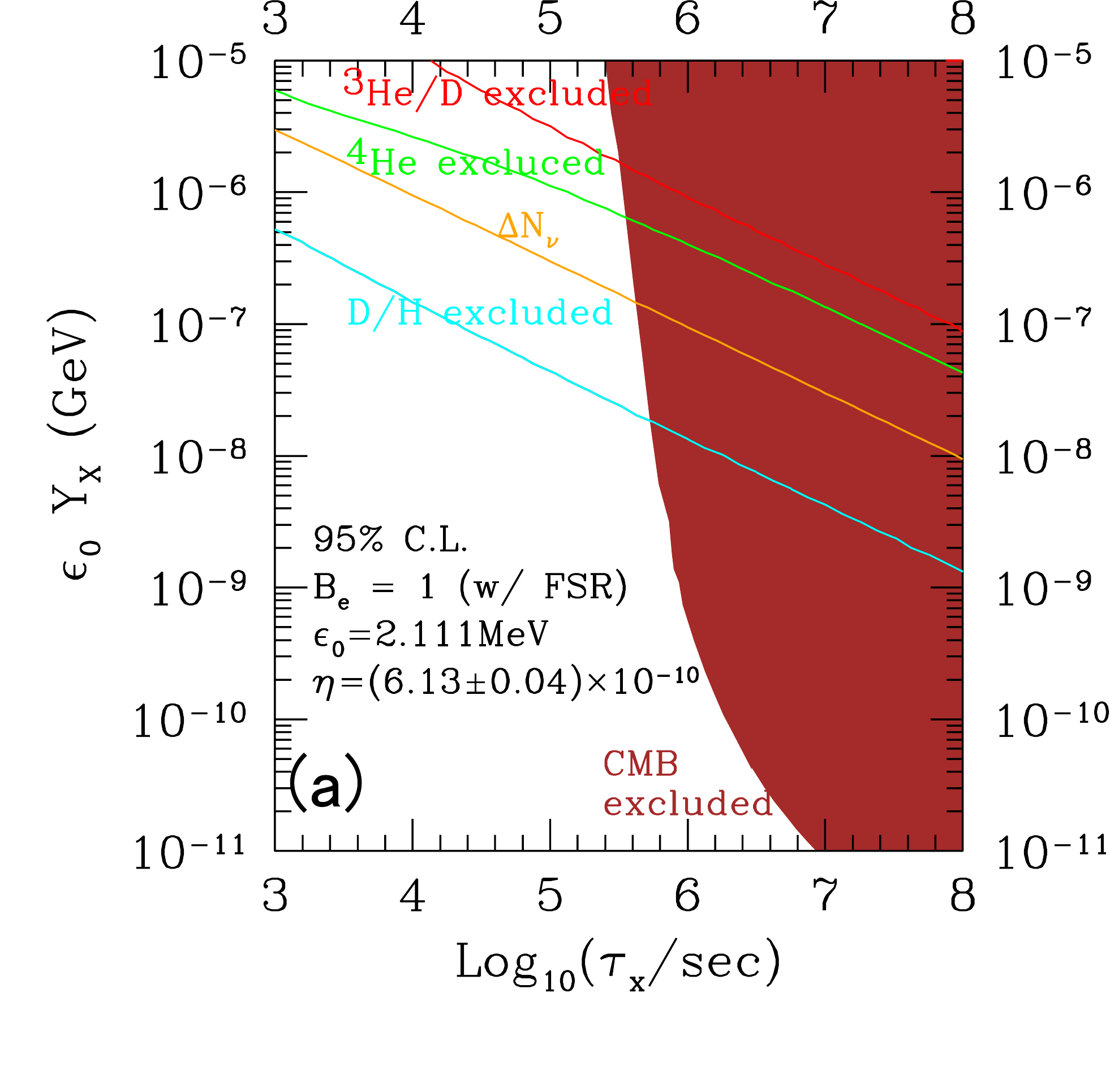}
\includegraphics[scale=0.36]{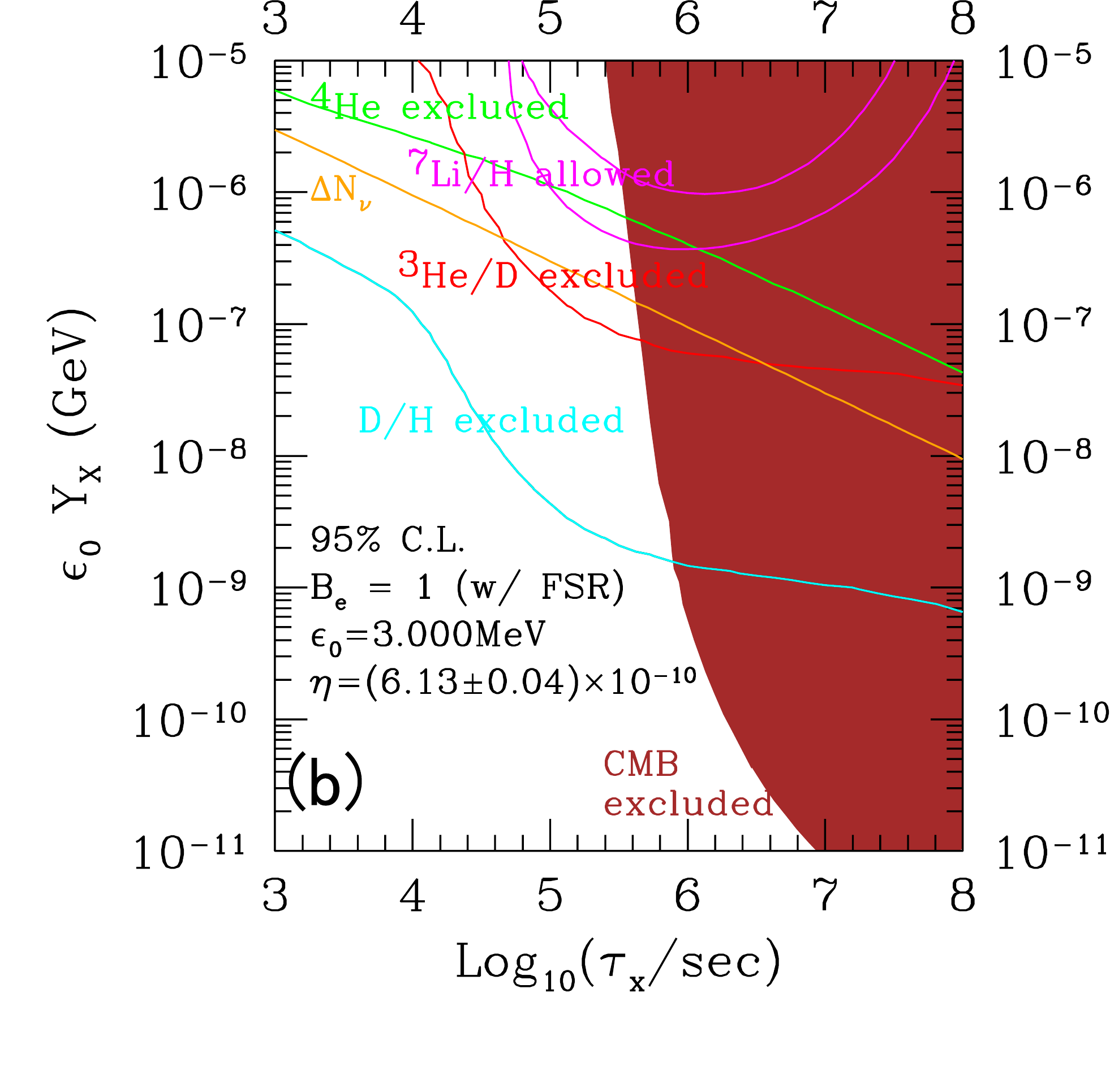}

%\Vspace{-1.0cm}
\caption{Allowed regions on $\epsilon_0 Y_X$ as a function of lifetime
  ($\tau_X$) to fit $^7$Li/H (magenta) for the energy of a high-energy
  injected electron, (a) $\epsilon_0 =2.1~{\rm MeV}$, and (b) $3.0~{\rm
    MeV}$ including effects of the FSR associated with electrons in
  the final states.  We executed Monte Carlo runs to obtain
  theoretical errors of light element abundances. The lines denote the
  bounds at 95$\%$ C.L. from $^3$He/D (red), Y$_p$ (green), D/H (cyan), and the $\mu$-  and y-distortion of CMB (brown), respectively.}
\label{fig:myxGMeVefsr}
\end{center}
\end{figure}
%%%%%%%%%%%%%%%%%%%%%%%%%%%%%%%%%%%%%%%%%%%%%%%%%%%%%%%%%%%%%%%%%%%%%%

%%%%%%%%%%%%%%%%%%%%%%%%%%%%%%%%%%%%%%%%%%%%%%%%%%%%%%%%%%%%%%%%%%%%%%
\section{Conclusions and Discussion}
\label{sec:conclusions}
%%%%%%%%%%%%%%%%%%%%%%%%%%%%%%%%%%%%%%%%%%%%%%%%%%%%%%%%%%%%%%%%%%%%%%

In this paper, we have studied the effects of the injections of
energetic electromagnetic particles (i.e., $\gamma$ and $e^\pm$),
paying particular attention to the case that the injection energy is
sub-GeV.  Once the energetic electromagnetic particles are injected
into the thermal bath in the early Universe, they induce
electromagnetic showers in which energetic photons are copiously
produced.  If it happens at a cosmic time later than
$\sim 1\ {\rm sec}$, such energetic photons in the shower may
dissociate the light elements (i.e., D, $^3$He, $^4$He, and so on),
resulting in the change of the predictions of the SBBN.  Because the
predictions of the SBBN more or less agree with the observations of
the primordial abundances of light elements, the injections of the
energetic particles in such an epoch is dangerous, and we can obtain
an upper bound on the total amount of the injection in order not to spoil the success of the SBBN.

We have concentrated on a long-lived particle $X$ which decays into a
pair of photon or $e^\pm$, and derived an upper bound on its
primordial abundance (Thus, the injection energy is monochromatic if
we neglect the FSR).  When the injection energy is higher than
$\sim 1\ {\rm GeV}$, it has been known that the resultant photon and
$e^\pm$ spectra in the electromagnetic shower are mostly determined by
the total amount of energy injection.  For a lower injection energy,
on the contrary, the spectra depends on the primary particle injected
and the injection energy, as was pointed out by
\cite{Poulin:2015woa,Poulin:2015opa,Forestell:2018txr}.

In our analyses, we have first solved the Boltzmann equations to
derive the distributions of photon and $e^\pm$ in the electromagnetic
shower.  The photon spectrum is convoluted with the photodissociation
cross sections of light elements to calculate the dissociation rates.
Effects of the photodissociations, based on the rates mentioned above,
are implemented into a numerical code to follow the evolutions of the
light elements with taking into account the effects of
photodissociations induced by the injection of energetic $\gamma$ or
$e^\pm$.  The theoretical predictions about the light element abundances
are compared with the latest observational constraints to derive
upper bound on the primordial abundance of the long-lived particle.

When the injection energy $\epsilon_0$ is high enough
($\epsilon_0\gtrsim 1\ {\rm GeV}$), the upper bound on the combination
of $\epsilon_0Y_X$ is insensitive to the primary particle injected and
the injection energy.  For smaller injection energy (i.e.,
$\epsilon_0\lesssim 100\ {\rm MeV}$), on the contrary, the upper
bounds become dependent on those injection energies.  For smaller
$\epsilon_0$, the upper bounds on $\epsilon_0Y_X$ is weaker than those
for $\epsilon_0\gtrsim 1\ {\rm GeV}$; this is because, for small
$\epsilon_0$, energy transfer from the energetic particle to the
scattered particle (which originally belongs to the thermal bath)
becomes inefficient.

We have also discussed the effects of the injection of the
electromagnetic particles on the $^7$Li abundance and discussed the
implication to the so-called $^7$Li problem.  If we regard the
Spike-plateau value of the $^7$Li abundance as the primordial value, the
SBBN overproduces the $^7$Li abundance by the factor of
$\sim 3$.  If the energetic particle injected into the thermal bath
can selectively dissociate $^7$Be, which is the dominant source of
$^7$Li for the value of the baryon-to-photon ratio observed by CMB,
the resultant abundance of $^7$Li can be reduced to be consistent with
the Spike-plateau value.  In Fig.7 (c), we have shown that there
really exists a parameter region in which the theoretical prediction
of the $^7$Li abundance becomes consistent with the Spike-plateau
value without conflicting other constraints; this is particularly due
to the smallness of the threshold energy of the $^7$Be.

In this paper, we have performed a general analysis to study the
effects of the injections of sub-GeV electromagnetic on the BBN.  In
particular, we have considered the simplest case that the injection
energy is assumed to be monochromatic, treating the lifetime and
injection energy (or the mass of the $X$) to be free parameters.  Our
analysis can be easily applied to models containing a long-lived
unstable particle with its mass of sub-GeV.  In particular, in some
class of models of sub-GeV dark matter, this is the case.  One example
is the so-called Twin-SIMPs model, in which strongly interacting
massive particles (SIMPs) provide a dark matter candidate
\cite{Hochberg:2018vdo}.  In this model, there show up various bound
states, whose masses are sub-GeV, due to the newly introduced QCD-like
strong interaction.  The lightest ``meson'' becomes the dark matter
candidate, while other ``mesons'' may decay into standard model
particles with lifetime longer than $1\ {\rm sec}$.  Because the
``mesons'' are expected to be produced in the thermal bath in the
early Universe, some of them may play the role of $X$ in our
discussion and affect the light element abundances.  In the Twin-SIMPs
model, the unstable ``mesons'' may decay into three or more final
states so that the decay products are not monochromatic in general.
Thus a dedicated analysis for the Twin-SIMPs model is required to
understand possible BBN constraints on the model.  Such a study is
beyond the scope of this paper, and will be given elsewhere
\cite{KKMMM_Future}.

%%%%%%%%%%%%%%%%%%%%%%%%%%%%%%%%%%%%%%%%%%%%%%%%%%%%%%%%%%%%%%%%%%%%%%
\section*{Acknowledgments}
%%%%%%%%%%%%%%%%%%%%%%%%%%%%%%%%%%%%%%%%%%%%%%%%%%%%%%%%%%%%%%%%%%%%%%
This work is supported in part by JSPS KAKENHI grant Nos.\ JP17H01131
(MK and KK), 17K05434 (MK), 16H06490 (TM), 18K03608 (TM), MEXT KAKENHI Grant
Nos.\ 15H05889 (MK and KK), 19H05114 (KK), 20H04750 (KK), 17K05409 (HM), World Premier International Research Center Initiative (WPI Initiative), MEXT, Japan (MK, KK, TM, KM and HM), the Program of Excellence in Photon Science (KM), MEXT Grant-in-Aid for Scientific Research on Innovative Areas JP15H05887, JP15K21733 (HM), the JSPS Research Fellowships for Young Scientists Grant No. 20J20248 (KM), and NSF grant PHY-1915314 and U.S. DOE Contract DE-AC02-05CH11231 (HM).  HM is also supported by Hamamatsu Photonics K.K. as Hamamatsu Professor.

%%%%%%%%%%%%%%%%%%%%%%%%%%%%%%%%%%%%%%%%%%%%%%%%%%%%%%%%%%%%%%%%%%%%%%
\appendix
%%%%%%%%%%%%%%%%%%%%%%%%%%%%%%%%%%%%%%%%%%%%%%%%%%%%%%%%%%%%%%%%%%%%%%

%%%%%%%%%%%%%%%%%%%%%%%%%%%%%%%%%%%%%%%%%%%%%%%%%%%%%%%%%%%%%%%%%%%%%%
\section{Analytical Formula of the Dilution Factor}
\label{sec:analyticD}
%%%%%%%%%%%%%%%%%%%%%%%%%%%%%%%%%%%%%%%%%%%%%%%%%%%%%%%%%%%%%%%%%%%%%%
\setcounter{equation}{0}

Here we discuss the analytical formula of the dilution factor induced
by the late-time entropy production due to radiatively decaying
particles. For simplicity, we assume that the parent particle (called
$X$) decays into photons instantaneously with the increase of the
photon energy density $\Delta \rho_{\gamma}$ at a cosmic time
$t = \tau_{X}$ where $\tau_{X}$ is the lifetime of $X$. In addition,
we assume that the emitted photons are immediately thermalized to be a
black body distribution.\footnote
{When we seriously consider the
  $\mu$-distortion and $y$-distortion from the perfect black body
  distribution, the current simple picture is incorrect. However,
  outside the parameter regions which are excluded by the severe
  observational constraint from the $\mu$-distortion and
  $y$-distortion, it is reasonable to assume that the black body
  distribution is approximately established. }
Then, we can
approximately estimate the dilution factor as
\begin{eqnarray}
    \label{eq:dilutin_factor}
    \Delta_{\rm dilution} = 1 + \frac{\Delta s}{s} 
    = \left(\frac{T_{ a}}{T_{ b}} \right)^{3},
\end{eqnarray}
where $T_{ a}$ and $T_{ b}$ are the photon temperature just after and
just before the entropy production, respectively. Here $\Delta s$ is
the increase of the entropy density. On the other hand,  $\Delta
\rho_{\gamma}$ is related with $T_{ a}/T_{ b}$ by
\begin{eqnarray}
    \label{eq:detla_rho}
    \frac{ \rho_{\gamma} + \Delta \rho_{\gamma}}{\rho_{\gamma}} 
    = \left(\frac{T_{ a}}{T_{ b}} \right)^{4}.
\end{eqnarray}
From Eqs.~(\ref{eq:dilutin_factor}) and (\ref{eq:detla_rho}), we have
\begin{eqnarray}
    \label{eq:dilution_deltarho}
    \Delta_{\rm dilution} 
   = \left( 1 + \frac{\Delta \rho_{\gamma}}{\rho_{\gamma}} \right)^{3/4}.
\end{eqnarray}

Hereafter we consider only the entropy production after the
$e^{+}e^{-}$--annihilation epoch, i.e., $T \lesssim 1 {\rm MeV}$.
Although this assumption is not correct for shorter lifetime in
general, you will find later that it is reasonable in the parameter
regions for the current interests.  Using the relation
$s = 7.04 n_{\gamma}$ between the entropy density $s$ and the number
density of photon $n_{\gamma}$, we obtain
\begin{eqnarray}
    \label{eq:delrhorho}
    \frac{\Delta \rho_{\gamma}}{\rho_{\gamma}}
     = \frac{7.04}{2.701} \frac1T \frac{\Delta \rho_{\gamma}}{s},
\end{eqnarray}
where we used $\rho_{\gamma} = 2.701 T n_{\gamma}$.

From the Friedmann equation, the temperature at $t = \tau_{X}$  is
approximately expressed by
\begin{eqnarray}
    \label{eq:friedmann}
    T = 1.556 \times 10^{-3 } {\rm GeV} g_{*}^{-1/4}
    \left(\frac{\tau_{X}}{{\rm sec}}\right)^{-1/2},
\end{eqnarray}
with the statistical degree of freedom,
\begin{eqnarray}
    \label{eq:gstar}
    g_{*} = 3.363 + 2 \frac{\Delta \rho_{\gamma}}{\rho_{\gamma}}.
\end{eqnarray}

By substituting Eq.~(\ref{eq:friedmann}) into
Eq.~(\ref{eq:delrhorho}), we have a quartic equation for $x = \Delta
\rho_{\gamma}/{\rho_{\gamma}}$,
\begin{eqnarray}
    \label{eq:quartic_equation}
    A^{4} x^{4} - 2 C^{4} x - C^{4} B = 0,
\end{eqnarray}
with
\begin{eqnarray}
    \label{eq:ABC}
    A = 0.5970 \times 10^{-3} \left( \frac{\tau_{X}}{\rm sec}
    \right)^{-1/2}, \quad
    B = 3.363,  \quad
    C = \left( \frac{\Delta \rho_{\gamma}/s}{\rm GeV} \right).
\end{eqnarray}

The exact solution for a positive real number of  $x$ is analytically
represented by
\begin{eqnarray}
    \label{eq:solution}
    x = \Delta \rho_{\gamma}/{\rho_{\gamma}} = \frac{C}{2}
  \left( \sqrt{D} + \sqrt{ - D + \frac{4C}{A^{4}} \frac1{\sqrt{D}}  } \right)
\end{eqnarray}
where
\begin{eqnarray}
    \label{eq:DefD}
    D = 2^{1/3}  3^{-2/3} A^{-8/3}  F - 2^{5/3} 3^{-1/3} A^{-4/3} B / F,
\end{eqnarray}
with
\begin{eqnarray}
    \label{eq:DefF}
    F = \left(9 C^{2} + \sqrt{3} \sqrt{16A^{4} B^{3} + 27 C^{4}}
    \right)^{1/3}.
\end{eqnarray}
By using this solution, we can calculate $\Delta_{\rm dilution}$ through
Eq.~(\ref{eq:dilution_deltarho}).

\begin{figure}[t]
    \begin{center}
      \includegraphics[scale=0.36]{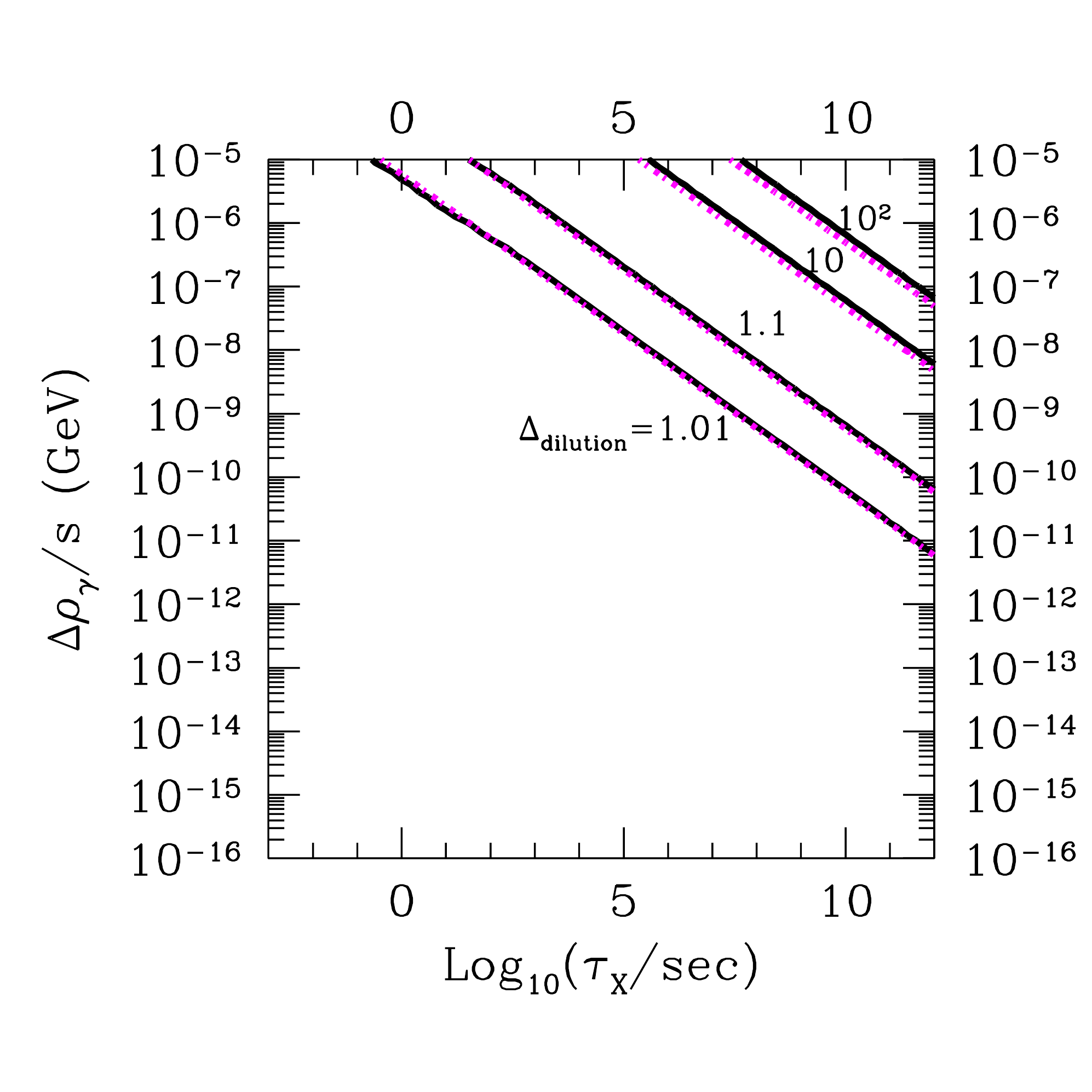}
        \caption{Contours of the dilution factor $\Delta_{\rm dilution}$ in the
        $\tau_{X}$ -- $\Delta  \rho_{\gamma}/s$ plane. The solid lines
        represent the results of the numerical computation. The dotted
        lines are the analytical solutions which are  given in
        Eqs.~(\ref{eq:dilution_deltarho}) and ~(\ref{eq:solution}).}
        \label{fig:fac_dilution}
    \end{center}
\end{figure}

In Fig.~\ref{fig:fac_dilution} we plot the contours of the dilution
factor $\Delta_{\rm dilution}$ in the $\tau_{X}$ -- $\Delta
\rho_{\gamma}/s$ plane. The solid lines are the results of the
numerical computation.\footnote
{Here the numerical results are
  obtained by numerically computing the entropy production due to the
  radiatively decaying $X$, which obeys the differential equation
  $dn_{X}/dt = -n_{X}/\tau_{X}$, in the expanding Universe.}
The dotted lines are the analytical formula given in
Eqs.~(\ref{eq:dilution_deltarho}) and~(\ref{eq:solution}). From this
figure, we see that the analytical formula fits the numerical results
very well. In this parameter space, the meaningful entropy production
occurs for $\tau_{X} \gtrsim 1 {\rm sec}$, which corresponds to the
decay epoch $T \lesssim 1 {\rm MeV}$. Therefore the assumption to
derive Eq.~(\ref{eq:solution}) that we considered only the
relationship among the physical variables after the
$e^{+}e^{-}$--annihilation is reasonable.

Because the exact solution  in Eq.~(\ref{eq:solution}) is little bit
complicated, it might be useful to give a simpler approximate
solution. When we join the solutions for both the limit cases of $C =
\Delta \rho_{\gamma}/s/{\rm GeV} \to \infty$ and  $C \to 0$, we obtain
\begin{eqnarray}
    \label{eq:simple}
       \Delta \rho_{\gamma}/\rho_{\gamma}
    = \left\{
        \begin{array}{ll}
            \displaystyle{
             \frac{B^{1/4}}{A} 
            C
            }
            &
            ~~:~~ 
            \displaystyle{ C \le  \frac12 A B^{3/4} }
            \\ \\
            \displaystyle{
             2^{1/3} \left(\frac{C}{A} \right)^{4/3}
            }
            &
            ~~:~~ 
            \displaystyle{ C > \frac12 A B^{3/4} }
        \end{array}
    \right..
\end{eqnarray}
This simple formula also agrees with the exact solution within the
precision of {\cal O}(1) $\%$ for
$\Delta_{\rm dilution} \ll {\cal O}(10)$.

%%%%%%%%%%%%%%%%%%%%%%%%%%%%%%%%%%%%%%%%%%%%%%%%%%%%%%%%%%%%%%%%%%%%%%%
\section{Final State Radiation}
\label{sec:FSR}
%%%%%%%%%%%%%%%%%%%%%%%%%%%%%%%%%%%%%%%%%%%%%%%%%%%%%%%%%%%%%%%%%%%%%%%

In this appendix, we derive the formulae for the rate of a decay process with a FSR photon under some assumptions.

\begin{figure}[t]
  \begin{center}
    \includegraphics[width=0.9\textwidth]{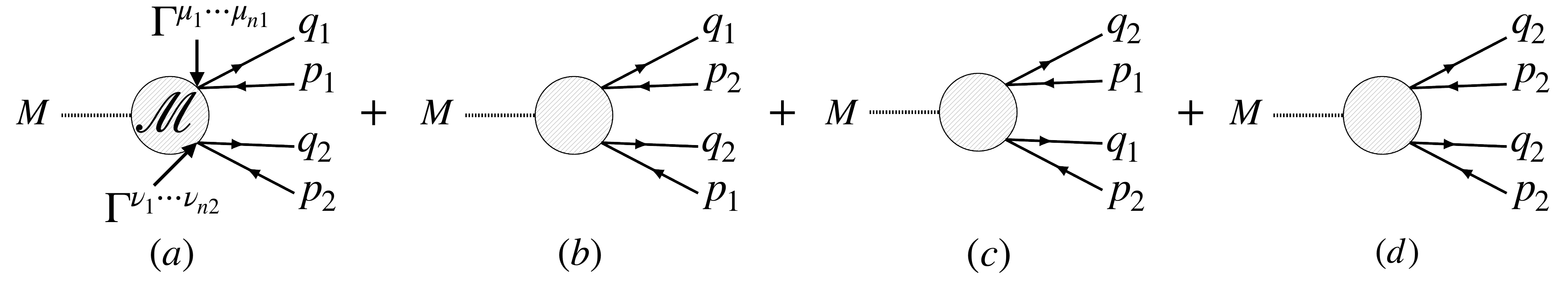}
    \caption{The diagrams without the FSR.  }
    \label{fg:14}
  \end{center}
%\end{figure}
%%
%%
%\begin{figure}[h]
  \begin{center}
    \includegraphics[width=0.9\textwidth]{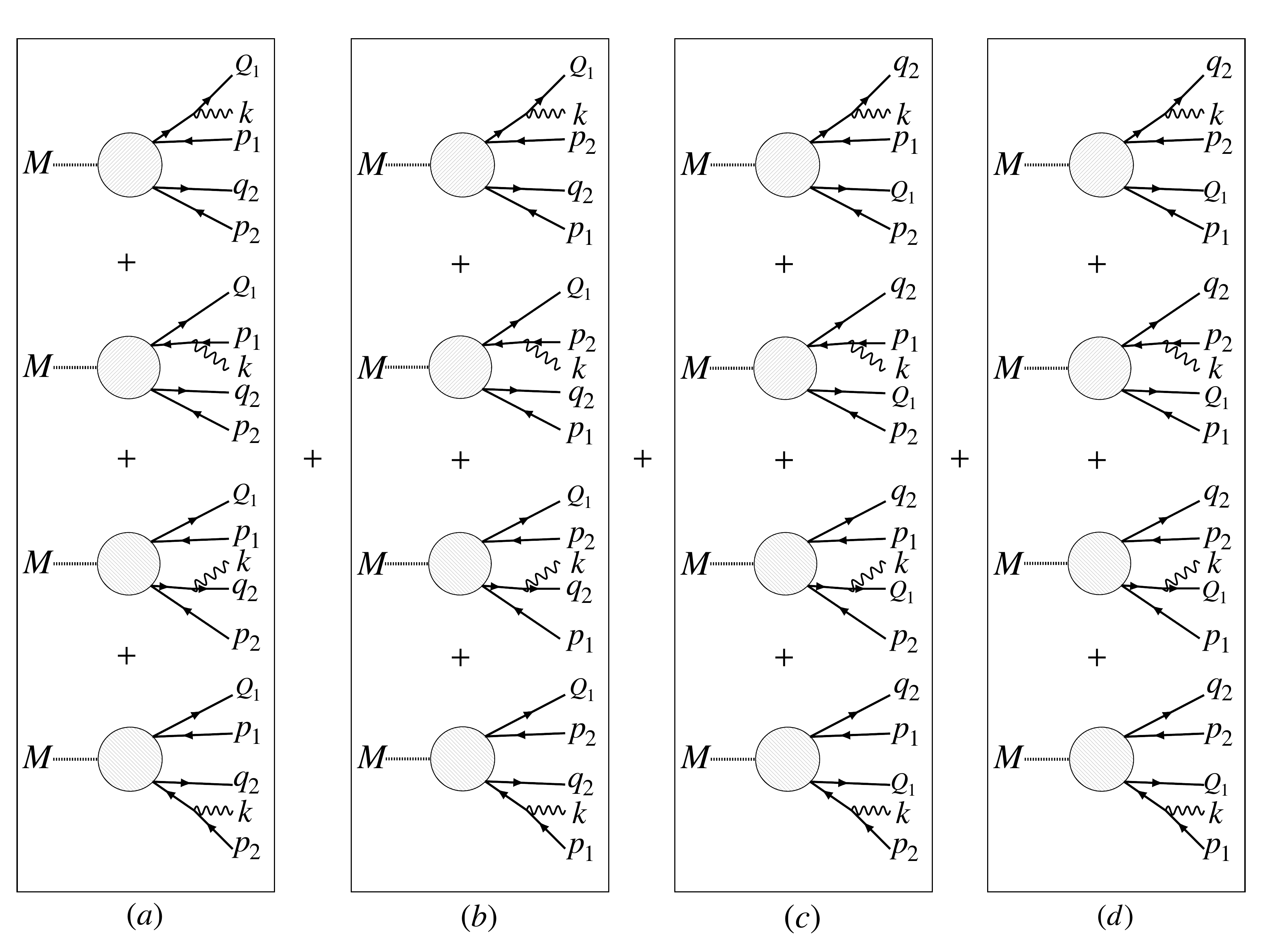}
    \caption{The diagram with the FSR. }
    \label{fg:15}
  \end{center}
\end{figure}

First, we consider the process in which two ultra-relativistic electron-positron pairs are emitted and compare the decay rate of the process with a FSR photon and that without FSR photons.
The diagrams to be considered are shown in Fig. \ref{fg:14}, \ref{fg:15}.
In Fig. \ref{fg:14}, \ref{fg:15}, $\Gamma_1^{\mu_1 \cdots \mu_{n_1}} \Gamma_2^{\nu_1 \cdots \nu_{n_2}}\mathcal{M}_{\mu_1 \cdots \mu_{n_1} \nu_1 \cdots \nu_{n_2}}$ represents the contribution from the shaded circle, and $\Gamma_1^{\mu_1 \cdots \mu_{n_1}}$ and $\Gamma_2^{\nu_1 \cdots \nu_{n_2}}$ represent products of gamma matrices which couple to one electron-positron pair and the other pair, respectively.
Here, we assume the momentum dependence of $\mathcal{M}$ is $\mathcal{M}_{\mu_1 \cdots \mu_{n_1} \nu_1 \cdots \nu_{n_2}}(q_1+p_1,q_2+p_2)$ for the diagram shown in Fig. \ref{fg:14}(a).
We assume the similar momentum dependences for the other diagrams in Fig.~\ref{fg:14}.
Under this assumption, the upper half diagrams in Fig. \ref{fg:15}(a) have $\mathcal{M}_{\mu_1 \cdots \mu_{n_1} \nu_1 \cdots \nu_{n_2}}(Q_1+p_1+k,q_2+p_2)$ and the lower half diagrams have $\mathcal{M}_{\mu_1 \cdots \mu_{n_1} \nu_1 \cdots \nu_{n_2}} (Q_1+p_1,q_2+p_2+k)$.
In addition, we assume each final state particle is ultra-relativistic.

Before calculating the decay rates, let us observe the diagrams with a FSR photon and determine which terms of the matrix element are dominant.
Comparing the diagrams with a FSR photon with those without FSR photons, the matrix element of the process with a FSR photon has one additional propagator of electron.
For example, Fig. \ref{fg:15}(a) has
\begin{equation}
  \frac{i(\cancel{Q}_1+\cancel{k}+m_e)}{(Q_1+k)^2-m_e^2} =
  \frac{i(\cancel{Q}_1+\cancel{k}+m_e)}{2k\cdot Q_1}.
\end{equation}
Since we assumed that each electron and positron is ultra-relativistic, the denominator is small when $k$ and $Q_1$ are collinear.
This is also the case with a FSR photon emitted from a positron.
Therefore, we can consider the cross section is dominated by the process in which the FSR photon is collinear with one of electrons or positrons.
Based on this observation, we proceed with the calculation
for the FSR photon collinear with $Q_1$.

In general, the decay rate of a massive particle with mass $M$ is written as
\begin{equation}
  \mathrm{d}\Gamma
  = \frac{1}{2M}
  \left(
   \prod_{\mathrm{f}}
   \frac{\mathrm{d}^3 p_\mathrm{f}}{(2 \pi)^32 p^0_{\mathrm{f}} }
   \right)
   \sum_{\mathrm{spin,polarization}}
   \Big|\mathcal{M}
    \left(
      M \to \{ p_\mathrm{f} \}
    \right)
   \Big|^2
   (2 \pi)^4 \delta^4
   \left(
    \sum_{\mathrm{f}} p_{\mathrm{f}} - M
   \right),
\end{equation}
where $p_{\mathrm{f}}$ is the momentum of final state particles
and we abbreviated $(M,\vec{0})$ as $M$.
In the case of Fig. \ref{fg:14}, the decay rate $\Gamma_{1 \to 4}$ is
\begin{align}
  \mathrm{d} \Gamma_{1\to 4}
  = &\frac{1}{2M}
  \frac{\mathrm{d}^3 q_1}{(2 \pi)^3 2 q_1^0}
  \frac{\mathrm{d}^3 q_2\mathrm{d}^3 p_1 \mathrm{d}^3 p_2}
    {8 (2 \pi)^{9} q_2^0 p_1^0 p_2^0}\nonumber\\
  &\times \sum_{\mathrm{spin}}
  \Big|
    \mathcal{M}_{1 \to 4} \left( q_1,q_2;p_1,p_2 \right)
  \Big|^2
  (2 \pi)^4 \delta^4
  \left( \sum_{\mathrm{f}} p_{\mathrm{f}} - M \right),
\end{align}
where the matrix element is $\mathcal{M}_{1\to4}$.
In the case of Fig. \ref{fg:15}, the decay rate $\Gamma_{1\to 5}$ is
\begin{align}
  \mathrm{d} \Gamma_{1 \to 5}
  = &\frac{1}{2 M}
  \frac{\mathrm{d}^3 k \mathrm{d}^3 Q_1}{(2 \pi)^6 2 k^0 2 Q_1^0}
  \frac{\mathrm{d}^3 q_2 \mathrm{d}^3 p_1\mathrm{d}^3 p_2}
    {8 (2 \pi)^{9} q_2^0 p_1^0 p_2^0}\nonumber\\
  &\times \sum_{\mathrm{spin,polar}}
  \Big|\mathcal{M}_{1\to5}\left(k;Q_1,q_2;p_1,p_2\right)\Big|^2
  (2\pi)^4\delta^4\left(\sum_{\mathrm{f}} p_{\mathrm{f}} - M \right),
\end{align}
where the matrix element is $\mathcal{M}_{1 \to 5}$.
Considering the electron is ultra-relativistic,
\begin{equation}
  \mathrm{d}^3 k \mathrm{d}^3 Q_1
  \simeq
  \mathrm{d}k^0 \mathrm{d}Q^0_1 \mathrm{d} \Omega_{k} \mathrm{d} \Omega_{Q_1}
  (k^0)^2 (Q_1^0)^2.
\end{equation}
From the above observation, we integrate in the interval,
\begin{equation}
  0 \leq \theta \leq \theta_0 \quad , \quad
  \frac{m}{q_1^0} \ll \theta_0 \ll 1,
  \label{eq:theta}
\end{equation}
where $\theta$ is the angle between $\vec{k}$ and $\vec{Q_1}$.
This interval is sufficiently collinear and contains the region in which the integrand is not negligible.
In the collinear region, we can perform a change of variables such as
\begin{equation}
  \begin{split}
    k^0 &= x q_1^0,\\
    Q_1^0 &= (1-x) q_1^0.
    \label{eq:vartra}
  \end{split}
\end{equation}
Therefore, we obtain
\begin{equation}
  \begin{split}
    k^{\mu} &= x q_1^0(1,\hat{k}),\\
    Q_1^{\mu} &= (1-x) q_1^0(1 , v \hat{Q}_1),
  \end{split}
\end{equation}
where, from the on-shell condition of the electron,
\begin{align}
  (Q_1)^2 = m_e^2 &= (1-x)^2\left( q_1^0 \right)^2 (1-v^2)\nonumber\\
  &\sim 2 (1-x)^2 \left( q_1^0 \right)^2 (1-v) \quad
  (\mathrm{except \, for} \, x \sim 1).
\end{align}
That is
\begin{equation}
  1-v \sim \frac{m_e^2}{2 (1-x)^2 \left( q^0_1 \right)^2} \quad
  (\mathrm{except \, for} \, x \sim 1).
  \label{eq:v_approx}
\end{equation}
For the change of variables (\ref{eq:vartra}),
\[
  \left|
    \begin{array}{rr}
      \frac{\partial{k^0}}{\partial x}
      & \frac{\partial{Q_1^0}}{\partial x} \\
      \frac{\partial{k^0}}{\partial q_1^0}
      & \frac{\partial{Q_1^0}}{\partial q_1^0}
    \end{array}
  \right|
  =
  \left|
    \begin{array}{rr}
      q_1^0 & -q_1^0 \\
      x & 1-x
    \end{array}
  \right|
  = q_1^0,
\]
which leads to
\begin{equation}
  \mathrm{d}k^0\mathrm{d}Q^0_1 = q_1^0\mathrm{d}x\mathrm{d}q_1^0.
  \label{eq:x-q}
\end{equation}
From Eq.~(\ref{eq:x-q})
\begin{align}
  \frac{\mathrm{d}^3 k \mathrm{d}^3 Q_1}{(2 \pi)^6 2 k^0 2 Q_1^0}
%  &=
%  \frac{\mathrm{d}k^0 \mathrm{d}Q_1^0
%      \mathrm{d}\Omega_{k} \mathrm{d}\Omega_{Q_1}}
%    {4(2\pi)^6}
%  k^0 Q_1^0\nonumber\\
%  &=
%  \frac{2 \pi \mathrm{d}x \mathrm{d}q_1^0
%      \mathrm{d}\Omega_{q_1} \mathrm{d}(\cos{\theta})}
%  {4(2 \pi)^6}
%  x (1-x) \left( q_1^0 \right)^3\nonumber\\
  &= \frac{\mathrm{d}^3 q_1}{(2 \pi)^3 2 q_1^0}
  \mathrm{d}x \mathrm{d}(\cos{\theta})
  \frac{x (1-x)}{8 \pi^2} \left( q_1^0 \right)^2,
  \label{eq:dkdQ}
\end{align}
where, in the second equality, we changed the variables from $\vec{k}$ and $\vec{Q_1}$ to $\vec{q_1}$ and the angle between $\vec{k}$ and $\vec{Q_1}$.
Here, we assumed the integrand does not depend on the azimuthal angle
between $\vec{k}$ and $\vec{Q_1}$.
As seen below, this assumption is justified in the case concerned.
Using Eq.~(\ref{eq:dkdQ}) we obtain
\begin{align}
  \mathrm{d}\Gamma_{1\to 5} =
  &\frac{1}{2M}\frac{\mathrm{d}^3q_1}{(2\pi)^32q_1^0}
  \frac{\mathrm{d}^3q_2\mathrm{d}^3p_1\mathrm{d}^3p_2}
    {8(2\pi)^{9}q_2^0p_1^0p_2^0}
  \mathrm{d}x\mathrm{d}(\cos{\theta})\frac{x(1-x)}{8\pi^2}
  \left( q_1^0 \right)^2\nonumber\\
  &\times\sum_{\mathrm{spin,polar}}
  \Big|
    \mathcal{M}_{1\to5}\left(k;Q_1,q_2;p_1,p_2\right)
  \Big|^2
  (2\pi)^4\delta^4\left(\sum_{\mathrm{f}} p_{\mathrm{f}}-M\right).
  \label{eq:momentum}
\end{align}\\

Corresponding to the four diagrams (a)-(d) in Fig.\ref{fg:14} $\mathcal{M}_{1\to 4}$ is written as
\begin{equation}
  \mathcal{M}_{1\to4}
     = \mathcal{M}_{1\to4}^{(a)} + \mathcal{M}_{1\to4}^{(b)}
     + \mathcal{M}_{1\to4}^{(c)} + \mathcal{M}_{1\to4}^{(d)}.
\end{equation}
$\mathcal{M}_{1\to5}$ is written in the same way.
$\mathcal{M}^{(i)}_{1\to 4} (i=a,\ldots ,d)$ are given by
\begin{align}
  i\mathcal{M}_{1\to4}^{(a)} &=
  i\mathcal{M}_{\mu\nu}\bar{u}^{s_1}(q_1)\Gamma^{\mu}_1v^{s'_1}(p_1)
  \bar{u}^{s_2}(q_2)\Gamma^{\nu}_2v^{s'_2}(p_2),\\
  i\mathcal{M}_{1\to4}^{(b)} &=
  i\mathcal{M}_{\mu\nu}\bar{u}^{s_1}(q_1)\Gamma^{\mu}_1v^{s'_2}(p_2)
  \bar{u}^{s_2}(q_2)\Gamma^{\nu}_2v^{s'_1}(p_1),\\
  i\mathcal{M}_{1\to4}^{(c)} &=
  i\mathcal{M}_{\mu\nu}\bar{u}^{s_2}(q_2)\Gamma^{\mu}_1v^{s'_1}(p_1)
  \bar{u}^{s_1}(q_1)\Gamma^{\nu}_2v^{s'_2}(p_2),\\
  i\mathcal{M}_{1\to4}^{(d)} &=
  i\mathcal{M}_{\mu\nu}\bar{u}^{s_2}(q_2)\Gamma^{\mu}_1v^{s'_2}(p_2)
  \bar{u}^{s_1}(q_1)\Gamma^{\nu}_2v^{s'_1}(p_1),
\end{align}
where we abbreviated $\mu_1,\cdots,\mu_{n_1},\nu_1,\cdots,\nu_{n_2}$
as $\mu,\nu$.  Each of $\mathcal{M}^{(i)}_{1\to5}$ has four terms due
to the four possible insertion of a FSR photon to the external lines
of the corresponding diagram in Fig.~\ref{fg:14}.  For example,
\begin{equation}
\begin{split}
  i\mathcal{M}_{1\to5}^{(a)}
  = i\mathcal{M}_{\mu\nu}e\varepsilon^*_{\alpha}(k)
  \biggl[
  &\bar{u}^{s_1}(Q_1)\gamma^{\alpha}
  \frac{\cancel{Q}_1+\cancel{k}}{2k\cdot Q_1}\Gamma^{\mu}_1v^{s'_1}(p_1)
  \bar{u}^{s_2}(q_2)\Gamma^{\nu}_2v^{s'_2}(p_2)  \\
  + &\bar{u}^{s_1}(Q_1)\Gamma^{\mu}_1
  \frac{-\cancel{p}_1-\cancel{k}}{2k\cdot p_1}\gamma^{\alpha}v^{s'_1}(p_1)
  \bar{u}^{s_2}(q_2)\Gamma^{\nu}_2v^{s'_2}(p_2)\\
  + &\bar{u}^{s_1}(Q_1)\Gamma^{\mu}_1v^{s'_1}(p_1)
  \bar{u}^{s_2}(q_2)\gamma^{\alpha}
  \frac{\cancel{q}_2+\cancel{k}}{2k\cdot q_1}\Gamma^{\nu}_2v^{s'_2}(p_2)\\
  +  &\bar{u}^{s_1}(Q_1)\Gamma^{\mu}_1v^{s'_1}(p_1)
  \bar{u}^{s_2}(q_2)\Gamma^{\nu}_2
  \frac{-\cancel{p}_2-\cancel{k}}{2k\cdot p_2}\gamma^{\alpha}v^{s'_2}(p_2)
  \biggr].
\end{split}
\end{equation}
In the following, first we show
\begin{equation}
   \sum_{\mathrm{spin,polar}}
   \mathcal{M}_{1\to5}^{(i)} \mathcal{M}_{1\to5}^{(j)*}
    = F(x,q^0_1,\theta,v) \sum_{\mathrm{spin}}
   \mathcal{M}_{1\to4}^{(i)} \mathcal{M}_{1\to4}^{(j)*}
   \label{eq:target}
\end{equation}
for all $i,j = a,b,c,d$ using a function $F(x,q^0_1,\theta,v)$.
We then can show
\begin{equation}
  \sum_{\mathrm{spin,polar}}| \mathcal{M}_{1\to5}|^2
  = F(x,q^0_1,\theta,v) \sum_{\mathrm{spin}} |\mathcal{M}_{1\to4}|^2.
\end{equation}\\
Since $\mathcal{M}_{1\to5}^{(i)}$ and $\mathcal{M}_{1\to5}^{(j)*}$ have
four terms, $\mathcal{M}_{1\to5}^{(i)} \mathcal{M}_{1\to5}^{(j)*}$ has 16 terms.
But, as mentioned above, the dominant contribution comes from the term which has $k \cdot Q_1$ in the denominator.
Therefore, we focus on the 1 term with the product of the propagators
proportional to $\propto (k\cdot Q_1)^{-2}$ and the 6 terms with the product of the propagators proportional to
$\propto (k\cdot Q_1)^{-1}$.
$\sum_{\mathrm{spin}}\mathcal{M}_{1\to4}^{(i)}\mathcal{M}_{1\to4}^{(j)*}$ can be written as
\begin{equation}
  \sum_{\mathrm{spin}}\mathcal{M}_{1\to4}^{(i)}\mathcal{M}_{1\to4}^{(j)*}
  = \mathcal{M}_{\mu_1\nu_1}\mathcal{M}^*_{\mu_2\nu_2}
  \mathrm{tr}\left[\cdots \cancel{q}_1  \cdots \right].
\label{eq:-2start}
\end{equation}
Strictly, the RHS is the product of two traces for some $(i,j)$s, but here we ignored the difference because it does not affect the result.\\
Corresponding to this expression, the term proportional to
$\propto (k\cdot Q_1)^{-2}$ in
$\sum \mathcal{M}_{1\to5}^{(i)}\mathcal{M}_{1\to5}^{(j)*}$ is
\begin{equation}
  \sum_{\mathrm{spin,polar}}
  \mathcal{M}_{1\to5}^{(i)} \mathcal{M}_{1\to5}^{(j)*}
  \supset -e^2 \mathcal{M}_{\mu_1\mu_2} \mathcal{M}^*_{\mu_3\mu_4}
  \mathrm{tr}
  \left[ \cdots
  \frac{\cancel{Q}_1+\cancel{k}}{2k\cdot Q_1}
  \gamma^{\alpha}\cancel{Q}_1 \gamma_{\alpha}
  \frac{\cancel{Q}_1+\cancel{k}}{2k\cdot Q_1}
  \cdots \right],
\label{eq:-2totyuu}
\end{equation}
where we used $\sum_{\mathrm{polar}}\varepsilon_{\mu}(k)\varepsilon^*_{\nu}(k) = -g_{\mu\nu }$ and ignored $m_e$ in the propagators because of the ultra-relativisticity of electrons and positrons.
In addition, the part denoted by $\cdots$ in the trace is the same in Eqs.~(\ref{eq:-2start}) and (\ref{eq:-2totyuu}).
For the trace part of Eq.~(\ref{eq:-2totyuu}) notice that
\begin{align}
  \frac{\cancel{Q}_1+\cancel{k}}{2k\cdot Q_1}
  \gamma^{\alpha} \cancel{Q}_1 \gamma_{\alpha}
  \frac{\cancel{Q}_1+\cancel{k}}{2k\cdot Q_1}
%  &=-\frac{1}{2\left( k\cdot Q_1 \right)^2}
%  \left( \cancel{Q}_1+\cancel{k} \right) \cancel{Q}_1
%  \left( \cancel{Q}_1+\cancel{k} \right)\nonumber\\
%  &=-\frac{1}{2\left( k\cdot Q_1 \right)^2}\cancel{k}
%  \cancel{Q}_1 \cancel{k}\nonumber\\
%  &=-\frac{1}{2\left( k\cdot Q_1 \right)^2}
%  \left(2k \cdot Q_1 - \cancel{Q}_1 \cancel{k} \right)
%  \cancel{k}\nonumber\\
%  &=-\frac{\cancel{k}}{k \cdot Q_1}\nonumber\\
  &= -\frac{x \cancel{q}_1}{k \cdot Q_1},
\end{align}
where we have
%used $\gamma^{\mu}\gamma^{\nu}=2g^{\mu\nu}-\gamma^{\nu}\gamma^{\mu}$ and $\gamma^{\alpha}\gamma_{\mu}\gamma_{\alpha}=-2\gamma_{\mu}$, $\cancel{p}\cancel{p}=p^2$, and
ignored $ m_e^2 $.\\ Therefore, the contribution from the term
proportional to $\propto (k\cdot Q_1)^{-2}$ is
\begin{equation}
  \sum_{\mathrm{spin,polar}}
  \mathcal{M}_{1\to5}^{(i)} \mathcal{M}_{1\to5}^{(j)*}
  \supset
  \frac{e^2x}{k\cdot Q_1} \sum_{\mathrm{spin}}
  \mathcal{M}_{1\to4}^{(i)} \mathcal{M}_{1\to4}^{(j)*}.
\end{equation}\\

Next, we consider the 6 terms proportional to $\propto (k\cdot Q_1)^{-1}$.
For example, we focus on the terms whose propagators are proportional to $\propto (k\cdot p_1 k\cdot Q_1)^{-1}$.\\
First, notice that $\sum_{\mathrm{spin}}\mathcal{M}_{1\to4}^{(i)}\mathcal{M}_{1\to4}^{(j)*}$ is written in the following form:
\begin{equation}
  \sum_{\mathrm{spin}}
  \mathcal{M}_{1\to4}^{(i)} \mathcal{M}_{1\to4}^{(j)*}
  = \mathcal{M}_{\mu_1\mu_2} \mathcal{M}^*_{\mu_3\mu_4} \mathrm{tr}
  \left[
  \cdots \cancel{q}_1  \cdots \cancel{p}_1 \cdots
  \right].
  \label{eq:-1first}
\end{equation}
Corresponding to this expression, the term proportional to
$\propto (k\cdot p_1 k\cdot Q_1)^{-1}$ in
$\sum \mathcal{M}_{1\to5}^{(i)}\mathcal{M}_{1\to5}^{(j)*}$ is written as
\begin{equation}
  \begin{split}
    \sum_{\mathrm{spin,polar}}
    \mathcal{M}_{1\to5}^{(i)} \mathcal{M}_{1\to5}^{(j)*}
    \supset -e^2 \mathcal{M}_{\mu_1\mu_2} \mathcal{M}^*_{\mu_3\mu_4}
    \mathrm{tr}
    \biggl[
    &\cdots \cancel{Q}_1 \gamma^{\alpha}
    \frac{\cancel{Q}_1+\cancel{k}}{2k\cdot Q_1} \cdots \cancel{p}_1
    \gamma_{\alpha} \frac{-\cancel{p}_1-\cancel{k}}{2k\cdot p_1} \cdots \\
    +&\cdots \frac{\cancel{Q}_1+\cancel{k}}{2k\cdot Q_1}
    \gamma^{\alpha} \cancel{Q}_1 \cdots
    \frac{-\cancel{p}_1-\cancel{k}}{2k\cdot p_1} \gamma_{\alpha}
    \cancel{p}_1 \cdots
    \biggr].
  \end{split}
  \label{eq:-1trace}
\end{equation}
In order to keep the leading order,
we can ignore $k\cdot Q_1$ in the numerator.
In other words, we can assume $k\parallel Q_1$.
Therefore, the first term in the trace of (\ref{eq:-1trace}) is written as
\begin{align}
  \cdots \cancel{Q}_1 \gamma^{\alpha}
  \frac{\cancel{Q}_1+\cancel{k}}{2k\cdot Q_1} \cdots \cancel{p}_1
  \gamma_{\alpha}\frac{-\cancel{p}_1-\cancel{k}}{2k\cdot p_1}
  \cdots
%  \nonumber\\
%  =& \frac{-1}{4k\cdot Q_1k\cdot p_1} \cdots \cancel{Q}_1
%  \gamma^{\alpha}\left(\cancel{Q}_1+\cancel{k}\right) \cdots \cancel{p}_1
%  \gamma_{\alpha}\left(\cancel{p}_1+\cancel{k}\right) \cdots \nonumber\\
%  =& \frac{-1}{4k\cdot Q_1k\cdot p_1}
%  \cdots \cancel{Q}_1
%  \left(
%  2\left( Q_1^{\alpha}+k^{\alpha} \right)
%  - \left( \cancel{Q}_1+\cancel{k}\right)\gamma^{\alpha}
%  \right)
%  \cdots \cancel{p}_1 \gamma_{\alpha}
%  \left( \cancel{p}_1+\cancel{k} \right) \cdots \nonumber\\
%  =& \frac{-1}{2k\cdot Q_1k\cdot p_1}
%  \cdots \cancel{Q}_1 \cdots \cancel{p}_1
%  \left( \cancel{Q}_1+\cancel{k} \right)
%  \left( \cancel{p}_1+\cancel{k} \right)
%  \cdots \nonumber\\
%  =& \frac{-1}{2k\cdot Q_1k\cdot p_1}
%  \cdots \cancel{Q}_1  \cdots \cancel{p}_1
%  \left( \cancel{Q}_1+\cancel{k} \right)
%  \cancel{p}_1 \cdots \nonumber\\
%  =& \frac{-1}{2k\cdot Q_1k\cdot p_1}
%  \cdots \cancel{Q}_1 \cdots \cancel{p}_1
%  \left(
%  2p_1 \cdot (Q_1+k)-\cancel{p_1} \left( \cancel{Q}_1+\cancel{k} \right)
%  \right) \cdots\nonumber\\
%  =& \frac{-p_1 \cdot (Q_1+k)}{k\cdot Q_1k\cdot p_1}
%  \cdots \cancel{Q}_1 \cdots \cancel{p}_1 \cdots \nonumber\\
  =& \frac{-(1-x)}{xk\cdot Q_1}
  \cdots \cancel{q}_1 \cdots \cancel{p}_1 \cdots.
\end{align}
The second term in the trace of (\ref{eq:-1trace}) can be transformed in the same way.
As a result, the sum of the terms whose propagators are proportional to $(k\cdot p_1 k\cdot Q_1)^{-1}$ is
\begin{equation}
  \sum_{\mathrm{spin,polar}}\mathcal{M}_{1\to5}^{(i)}\mathcal{M}_{1\to5}^{(j)*}
  \supset \frac{2e^2(1-x)}{xk\cdot Q_1}\sum_{\mathrm{spin}}\mathcal{M}_{1\to4}^{(i)}\mathcal{M}_{1\to4}^{(j)*}.
\end{equation}
The 2 terms proportional to $\propto (k\cdot p_2 k\cdot Q_1)^{-1}$ and the 2 terms proportional to $\propto (k\cdot q_2 k\cdot Q_1)^{-1}$
are opposite in sign because of the sign of the propagators.
So, they completely cancel.
Thus, the contribution from the 6 terms proportional to $\propto (k\cdot Q_1)^{-1}$ is given by
\begin{align}
  \sum_{\mathrm{spin,polar}}
  \mathcal{M}_{1\to5}^{(i)} \mathcal{M}_{1\to5}^{(j)*}
  &\supset (1+1-1)\frac{2e^2(1-x)}{xk\cdot Q_1}
  \sum_{\mathrm{spin}}
  \mathcal{M}_{1\to4}^{(i)} \mathcal{M}_{1\to4}^{(j)*} \nonumber\\
  &=\frac{2e^2(1-x)}{xk\cdot Q_1} \sum_{\mathrm{spin}}
  \mathcal{M}_{1\to4}^{(i)} \mathcal{M}_{1\to4}^{(j)*}.
  \label{eq:-1final}
\end{align}
Summing all the contributions, we obtain
\begin{align}
  \sum_{\mathrm{spin,polar}}
  \mathcal{M}_{1\to5}^{(i)} \mathcal{M}_{1\to5}^{(j)*}
  &=e^2\left( x + \frac{2(1-x)}{x} \right)
  \frac{1}{k\cdot Q_1}
  \sum_{\mathrm{spin}} \mathcal{M}_{1\to4}^{(i)} \mathcal{M}_{1\to4}^{(j)*}
  + \mathcal{O} \left( (k\cdot Q_1)^0 \right) \nonumber\\
  &\simeq e^2\frac{1+(1-x)^2}{x^2(1-x)}
  \frac{1}{\left(q_1^0\right)^2(1-v\cos{\theta})}
  \sum_{\mathrm{spin}} \mathcal{M}_{1\to4}^{(i)} \mathcal{M}_{1\to4}^{(j)*}.
\end{align}
This is exactly the form of (\ref{eq:target}) and we can read  $F(x,q_1^0,\theta,v)$ from this expression as
\begin{equation}
  \sum_{\mathrm{spin,polar}}
  \left| \mathcal{M}_{1\to5} \right|^2
  \simeq e^2 \frac{1+(1-x)^2}{x^2(1-x)}
  \frac{1}{\left(q_1^0\right)^2(1-v\cos{\theta})}
  \sum_{\mathrm{spin}} \left| \mathcal{M}_{1\to4} \right|^2.
\label{eq:amplitude}
\end{equation}
Using (\ref{eq:momentum}) and (\ref{eq:amplitude}),
let us compare the decay rate with the FSR and that without FSRs.
Note that, because $F(x,q_1^0,\theta,v)$ is independent of the angle between $\vec{k}$ and $\vec{Q_1}$, we can use (\ref{eq:momentum}).
When the FSR photon is radiated collinear with the electron $Q_1$,
\begin{align}
  \mathrm{d}\Gamma_{1\to 5} \simeq
  &\frac{1}{2M}\frac{\mathrm{d}^3q_1}{(2\pi)^32q_1^0}
  \frac{\mathrm{d}^3q_2\mathrm{d}^3p_1\mathrm{d}^3p_2}
    {8(2\pi)^{9}q_2^0p_1^0p_2^0}
  \mathrm{d}x \mathrm{d} (\cos{\theta})
  \frac{x(1-x)}{8\pi^2} \left( q_1^0 \right)^2 \nonumber\\
  &\times e^2\frac{1+(1-x)^2}{x^2(1-x)}
  \frac{1}{\left(q_1^0\right)^2(1-v\cos{\theta})}
  \sum_{\mathrm{spin}} \left| \mathcal{M}_{1\to4} \right|^2
  (2 \pi)^4 \delta^4 \left( \sum p \right) \nonumber\\
%  =&\mathrm{d}x \mathrm{d} (\cos{\theta})
%  \frac{\alpha}{2\pi}
%  \frac{1+(1-x)^2}{x}\frac{1}{1-v\cos{\theta}}
%  \mathrm{d} \Gamma_{1\to4} \nonumber\\
  =&\mathrm{d}x\frac{\alpha}{2\pi}
  \frac{1+(1-x)^2}{x} \log
  \left[ \frac{1-\cos{\theta_0}}{1-v} \right] \mathrm{d} \Gamma_{1\to4}.
  \label{eq:compare}
\end{align}
From (\ref{eq:v_approx}),
\begin{align}
  &\frac{\mathrm{d}\Gamma_{1\to 5}}{\mathrm{d}x}
  (xq_1;(1-x)q_1,q_2;p_1,p_2) \nonumber\\
  &\simeq \frac{\alpha}{2\pi}
  \frac{1+(1-x)^2}{x} \log
  \left[ \frac{2(1-x)^2(q_1^0)^2(1-\cos{\theta_0)}}{m_e^2} \right]
  \mathrm{d} \Gamma_{1\to4} (q_1,q_2;p_1,p_2)
  \label{eq:15result}
\end{align}
except for $x \sim 1$.
Here, the arguments of $\frac{\mathrm{d}\Gamma_{1\to 5}}{\mathrm{d}x}$
are the momenta of photon, electrons and positrons from left to right.

In the following part, we will extend this result to the more general case.
First, we consider the case where the FSR photon is collinear with a positron.
In this case, we only have to change the sign of the propagators, and because we have calculated the product of two propagators in
$\sum|\mathcal{M}_{1\to 5}|^2$, this does not change the result:
\begin{align}
  &\frac{\mathrm{d}\Gamma_{1\to 5}}{\mathrm{d}x}
  (xp_1;q_1,q_2;(1-x)p_1,p_2) \nonumber\\
  &\simeq \frac{\alpha}{2\pi}\frac{1+(1-x)^2}{x}
  \log\left[ \frac{2(1-x)^2(p_1^0)^2(1-\cos{\theta_0)}}{m_e^2} \right]
  \mathrm{d}\Gamma_{1\to4}(q_1,q_2;p_1,p_2).
\end{align}
Second, we consider the case where the final state includes different species of fermion-anti-fermion pairs.
In this case, we have to change the classification of the diagrams in Fig. \ref{fg:14} and Fig. \ref{fg:15}.
However, we have calculated each $\mathcal{M}^{(i)}$ and the same derivation is also applicable in this case.
Therefore, the resultant expression is given by (\ref{eq:v_approx}) with the electron mass replaced by that of an appropriate fermion.
Third, we consider the case where the final state has $n$ pairs of fermions.
In this case, we have to change the calculation from (\ref{eq:-1first}) to (\ref{eq:-1final}).
Now, we have to calculate $2(2n-1)$ terms instead of 6 terms.
But each 4 terms cancel in the same way as the above calculation and we have only the same contribution from 2 terms left.
After all, we get the result:
\begin{align}
  &\frac{\mathrm{d}\Gamma_{1\to 2n+1}}{\mathrm{d}x}
  (xq_1;(1-x)q_1,q_2,\cdots,q_n;p_1,\cdots,p_n) \nonumber\\
  &\simeq \frac{\alpha}{2\pi}\frac{1+(1-x)^2}{x}
  \log \left[ \frac{2(1-x)^2(q_1^0)^2(1-\cos{\theta_0)}}{m^2} \right]
  \mathrm{d}\Gamma_{1\to2n} (q_1,\cdots,q_n;p_1,\cdots,p_n),
\label{eq:summary1}
\end{align}
\begin{align}
  &\frac{\mathrm{d}\Gamma_{1\to 2n+1}}{\mathrm{d}x}
  (xp_1;q_1,\cdots,q_n;(1-x)p_1,p_2,\cdots,p_n) \nonumber\\
  &\simeq \frac{\alpha}{2\pi}\frac{1+(1-x)^2}{x}
  \log \left[ \frac{2(1-x)^2(p_1^0)^2(1-\cos{\theta_0)}}{m^2} \right]
  \mathrm{d}\Gamma_{1\to2n} (q_1,\cdots,q_n;p_1,\cdots,p_n).
\label{eq:summary2}
\end{align}
Finally, we comment on the case where a final state fermion is collinear with another final state fermion.
Above, we performed the integration about the direction of the FSR photon by dividing the domain of integration into the domains where the FSR photon is collinear with one of fermions.
Therefore, this division can overlap in this case.
But, when the process is sufficiently ultra-relativistic, as we can see from (\ref{eq:theta}), $\theta_0$ about each fermion becomes so small that the overlap does not occur.

%%%%%%%%%%%%%%%%%%%%%%%%%%%%%%%%%%%%%%%%%%%%%%%%%%%%%%%%%%%%%%%%%%%%%%

%%%%%%%%%%%%%%%%%%%%%%%%%%%%%%%%%%%%%%%%%%%%%%%%%%%%%%%%%%%%%%%%%%%%%%

%%%%%%%%%%%%%%%%%%%%%%%%%%%%%%%%%%%%%%%%%%%%%%%%%%%%%%%%%%%%%%%%%%%%%%
\end{document}